\documentclass[a4paper,conference]{IEEEtran}
\usepackage{xspace}
\usepackage{bm}
\usepackage{adjustbox}
\usepackage{tabularx}
\usepackage{color}
\def\xstrut{\rule{0pt}{2ex}}

\usepackage{cite}

\ifCLASSINFOpdf
\else
\fi
\newcommand\blfootnote[1]{%
  \begingroup
  \renewcommand\thefootnote{}\footnote{#1}%
  \addtocounter{footnote}{-1}%
  \endgroup
}

\usepackage{amsmath}
\usepackage{algorithmic}

\usepackage{array}

\ifCLASSOPTIONcompsoc
  \usepackage[caption=false,font=normalsize,labelfont=sf,textfont=sf]{subfig}
\else
  \usepackage[caption=false,font=footnotesize]{subfig}
\fi
\usepackage{url}

\makeatletter
\DeclareRobustCommand\onedot{\futurelet\@let@token\@onedot} 
\def\@onedot{\ifx\@let@token.\else.\null\fi\xspace}

\def\eg{\emph{e.g}\onedot}

\def\etc{\emph{etc}\onedot} 
 
\def\etal{\emph{et al}\onedot}
\makeatother

\begin{document}
%
\title{CURL: Neural Curve Layers for  Global Image Enhancement}



\author{
\IEEEauthorblockN{Sean Moran$^{\ast,1}$, Steven McDonagh$^{2}$, Gregory Slabaugh$^{\ast,3}$}
 \IEEEauthorblockA{$^{1}$JP Morgan Chase \& Co., $^{2}$Huawei Noah's Ark Lab, $^{3}$Queen Mary University of London\\
 {sean.j.moran@jpmchase.com, steven.mcdonagh@huawei.com, g.slabaugh@qmul.ac.uk}}\\
\IEEEauthorblockN{} 
}


\definecolor{stevencolor}{rgb}{0.1,0.1,0.8}
\newcommand{\steven}[1]{{\textcolor{stevencolor}{[SGM #1]}}}
\newcommand\sgm[1] {\steven{#1}}

\definecolor{seancolor}{rgb}{0.8,0.1,0.1}
\newcommand{\sean}[1]{{\textcolor{seancolor}{[SM #1]}}}
\newcommand\sm[1] {\sean{#1}}

\definecolor{gregcolor}{rgb}{0.8,0.8,0.1}
\newcommand{\greg}[1]{{\textcolor{gregcolor}{[GS #1]}}}
\newcommand\gs[1] {\greg{#1}}

\maketitle
\blfootnote{
*Work done while at Huawei Noah's Ark Lab.
}
\begin{abstract}
We  present  a  novel  approach  to  adjust  global  image  properties such as colour, saturation, and luminance using human-interpretable image enhancement curves, inspired by the Photoshop curves tool. Our method, dubbed neural \textbf{CUR}ve \textbf{L}ayers (CURL), is designed as a multi-colour space neural retouching block trained jointly in three different colour spaces (HSV, CIELab, RGB) guided by a novel multi-colour space loss. The curves are fully differentiable and are trained end-to-end for different computer vision problems including photo enhancement (RGB-to-RGB) and as part of the image signal processing pipeline for image formation (RAW-to-RGB). To demonstrate the effectiveness of CURL we combine this global image transformation block with a pixel-level (local) image multi-scale encoder-decoder backbone network. In an extensive experimental evaluation we show that CURL produces state-of-the-art image quality versus recently proposed deep learning approaches in both objective and perceptual metrics, setting new state-of-the-art performance on multiple public datasets. Our code is publicly available at: \url{https://github.com/sjmoran/CURL}.
\end{abstract}


%
\IEEEpeerreviewmaketitle

\section{Introduction}

Image quality is of fundamental importance in any imaging system, especially DSLR and smartphone cameras.  Modern smartphones for example, provide multiple camera sensors and sophisticated image signal processor (ISP) pipelines to produce images with good contrast, detail, colours, and dynamic range while at the same time mitigating against degradations.  However, despite great advances in imaging hardware and ISP pipelines, there remains substantial room for improvement. Even professional photographers using DSLR will spend significant time using photo editing software to produce a quality digital photograph.

This paper is inspired by the Photoshop curves tool\footnote{\url{https://www.cambridgeincolor.com/tutorials/photoshop-curves.htm}}, which professional photographers  use to modify global properties of an image through manual design of adjustment curves. However, most casual users lack artistic skills (or software) to retouch their photos in such a manner. In this context, several interesting research questions emerge: \emph{Is it possible to automatically estimate, and apply, adjustment curves through multiple colour spaces to improve image quality? What collection of curves are necessary? What order should image adjustments be applied?} This paper not only precisely poses these research questions, but answers them using new neural CURve Layers dubbed \emph{CURL}.  We arrange a collection of such curve adjustments in a defined sequence operating in multiple colour spaces, resulting in a \emph{CURL block}.

\begin{figure*}[t!]
\centering
\begin{tabular}{c@{}c@{}c@{}c@{}}
      \scalebox{0.85}{Input} &
      \scalebox{0.85}{DeepUPE (16.85 dB)} & 
     \scalebox{0.85}{\textbf{Ours (23.55 dB)} } &
      \scalebox{0.85}{Groundtruth} \\
     \includegraphics[width=0.245\linewidth,scale=0.30]{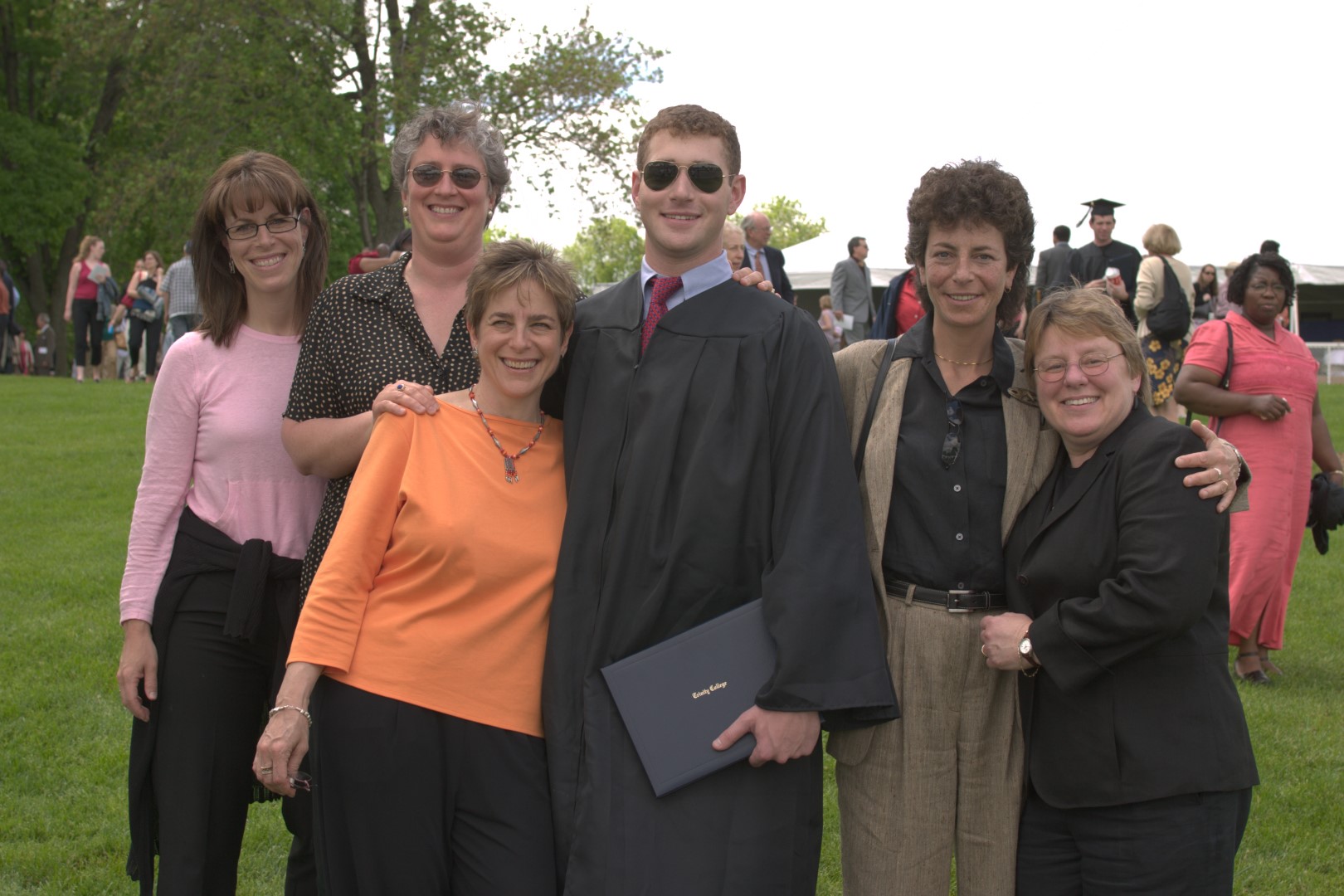} &
     \includegraphics[width=0.245\linewidth,scale=0.30]{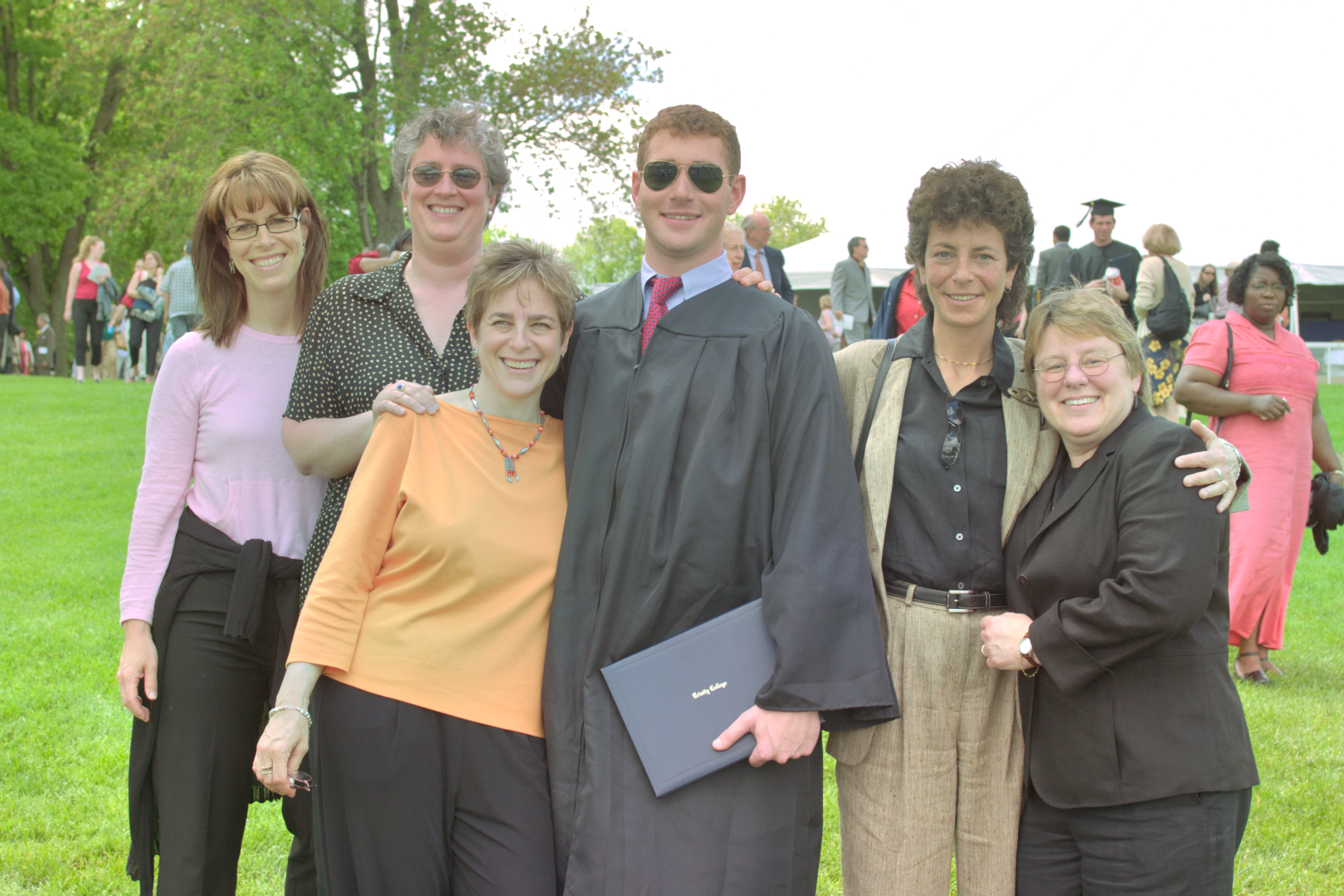} &
    \includegraphics[width=0.245\linewidth,scale=0.30]{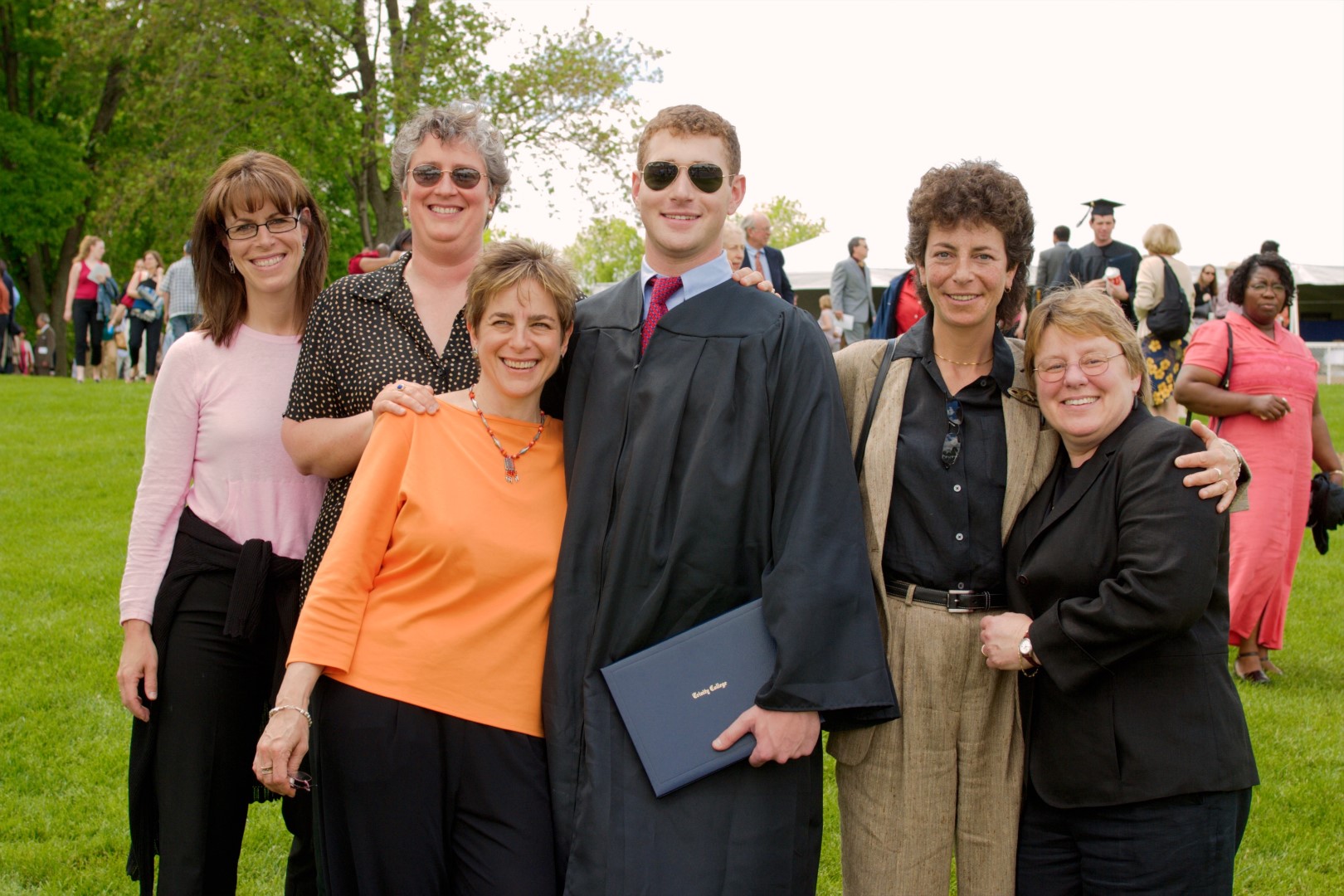} &
        \includegraphics[width=0.245\linewidth,scale=0.30]{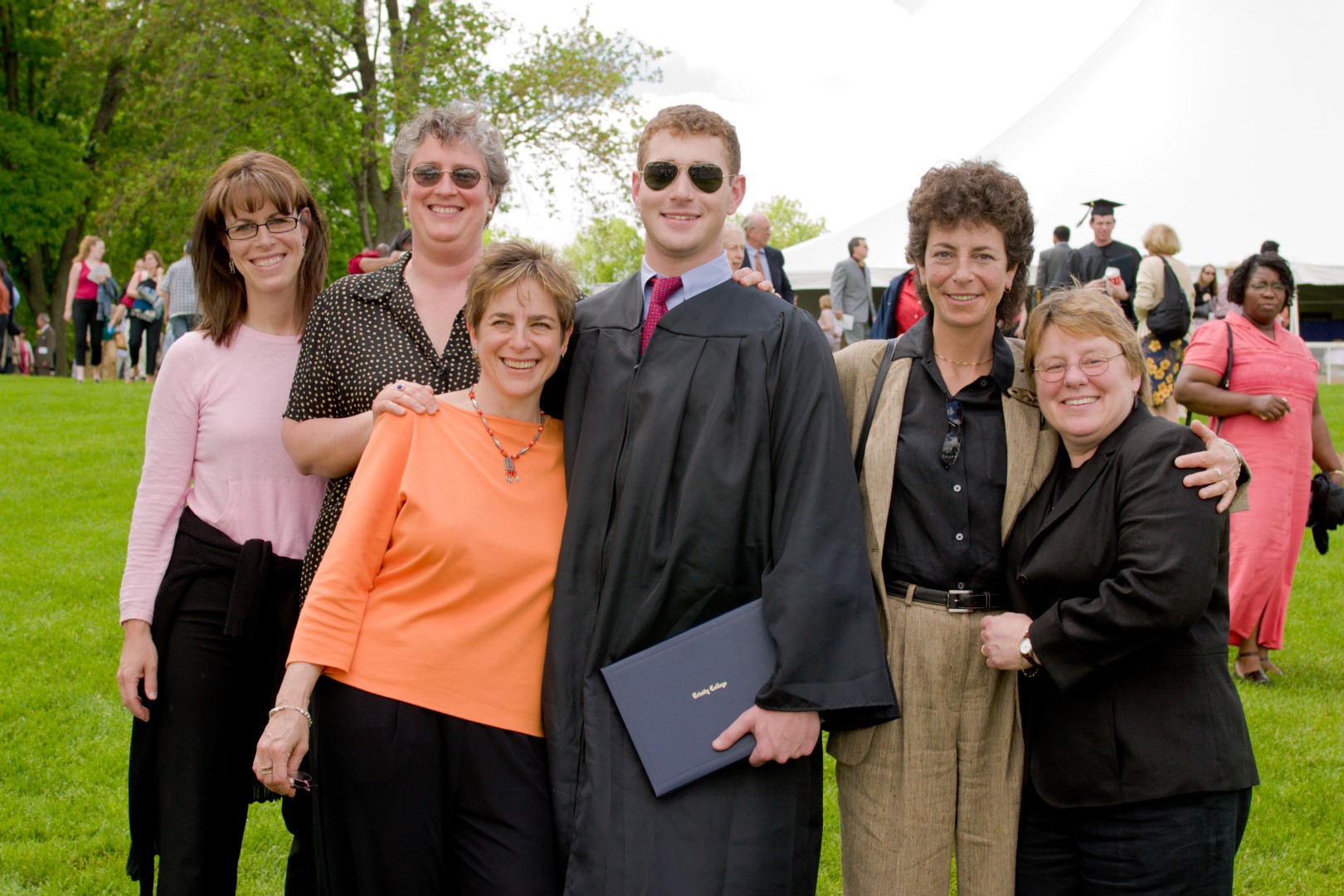}
\end{tabular}
    \caption{\textbf{Underexposed image enhancement}. Given a poorly exposed image, our method produces an image with pleasing contrast and colour better matching the groundtruth compared to the state-of-the-art DeepUPE model~\cite{wang19}.}
    \label{fig:example}
\end{figure*}

CURL can be used along with a deep backbone feature extractor to enhance images in an \textbf{RGB-to-RGB} mapping.  An example is shown in Figure~\ref{fig:example}, where CURL adjusts an underexposed image to produce a result with better contrast and more pleasing colours than other methods such as the state-of-the-art DeepUPE~\cite{wang19} approach.

Alternatively, a CURL block can be used as part of the \textbf{RAW-to-RGB} transformation modeling the entire ISP pipeline.  In this configuration, first a pixel level block (modeled as an encoder/decoder network) is applied to perform local operations such as denoising and demosaicing, the latter estimating RGB colours at each pixel from the RAW data collected using a colour filter array. Then, a CURL block is used to transform the brightness, colour, and saturation to produce the final image. Modeling the entire ISP in this way, as a single neural network trained end-to-end, is very appealing as standard ISP pipelines consist of many different modules (e.g. denoising, demosiacing, automatic white balance, tone mapping, etc.) that are sequentially applied. Such ISPs are difficult and expensive to design, develop, and tune due to the complex dependencies between the modules.

Our contributions in this paper are three-fold:

\begin{itemize}
    \item \textbf{Multi-colour space neural retouching block}: We introduce \textbf{CURL}, a neural retouching block, that learns a set of piece-wise linear scaling curves to globally adjust image properties in a human-interpretable manner, inspired by the Photoshop curves tool. 
    \item \textbf{Multi-colour space loss function}: We apply sequential differentiable transformations of the image in different colour spaces (Lab, RGB, HSV) guided by a novel multi-colour space loss function. 
    \item \textbf{Improved Encoder/Decoder for image enhancement}: CURL modifies images initially enhanced by a backbone network. We explore an effective backbone encoder/decoder architecture. Our \textbf{T}ransformed \textbf{E}ncoder-\textbf{D}ecoder backbone (dubbed \textbf{TED}) has all but one U-Net skip connection removed, with the remaining skip connection endowed with a multi-scale neural processing block that enriches the information available to the decoder. 
\end{itemize}

\section{Related Work}

\begin{figure*}[t!]
  \centering
  \includegraphics[scale=0.23]{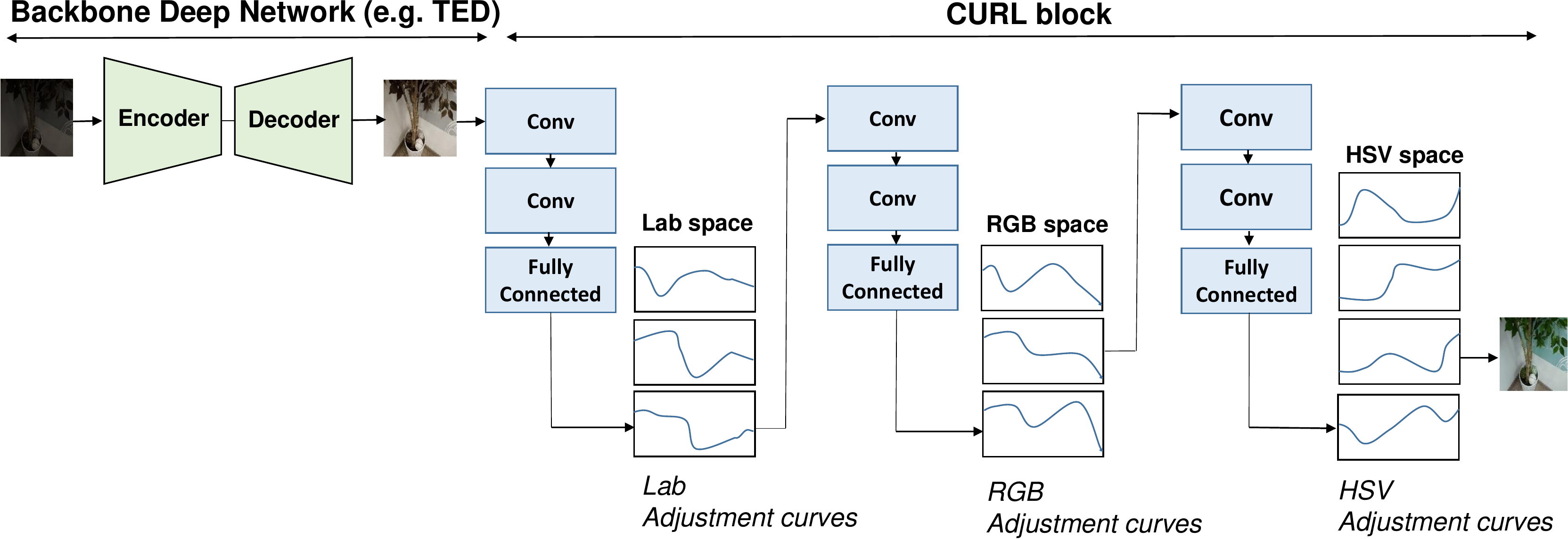}
  \includegraphics[scale=0.23]{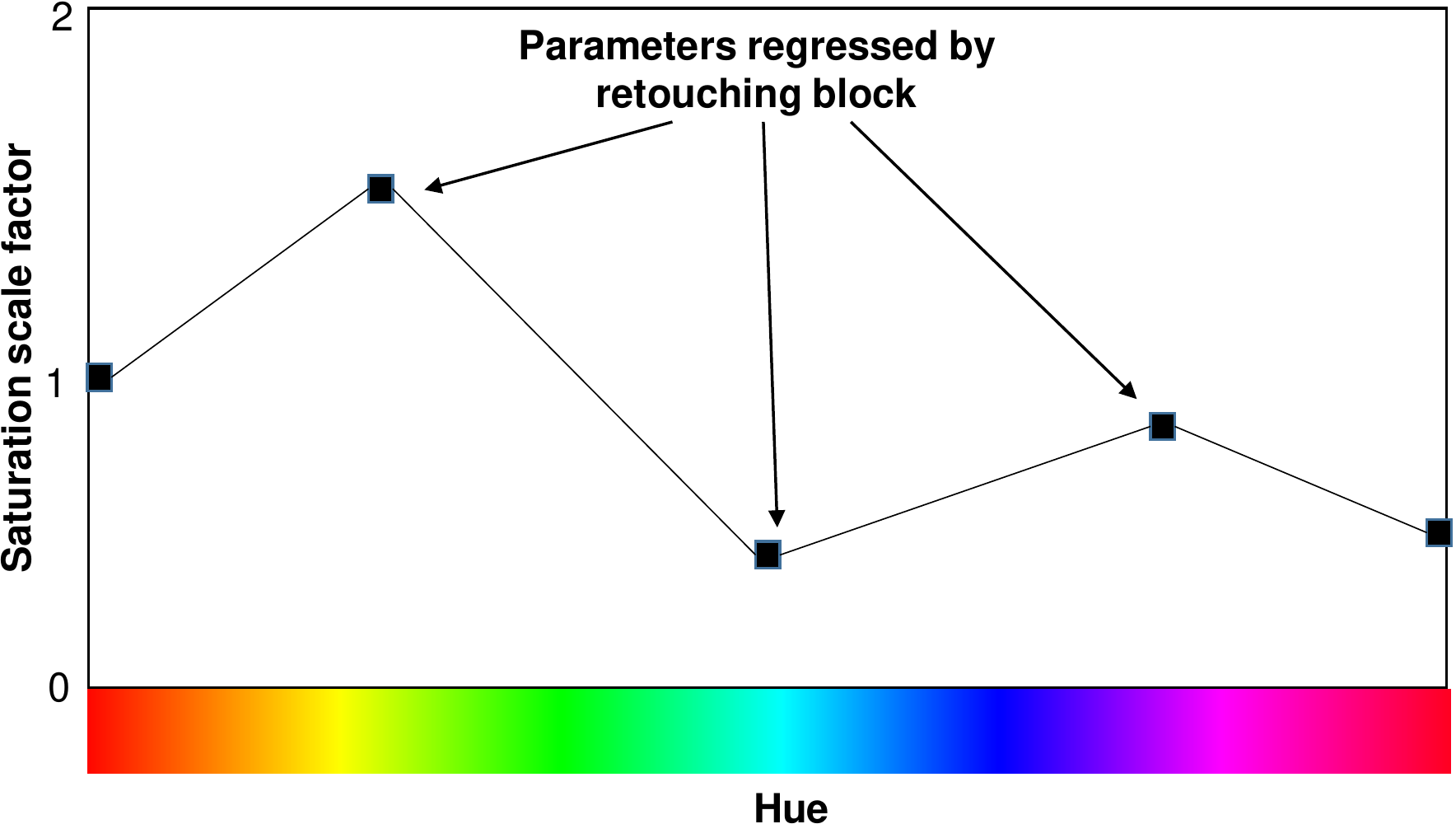}
  \caption{\textbf{Left:} Placement of the CURL block with regards to an encoder-decoder backbone. The CURL block leverages \emph{neural curve layers} for artist-inspired image refinement operations. The input to the backbone network is a RAW image and the output from the CURL block is a high quality RGB image with correct colour and brightness. \textbf{Right:} Illustration of a piece-wise linear neural curve from the CURL block that predicts a scaling factor that adjusts saturation based on hue.}
  \label{fig:DIFAR_pipeline}
\end{figure*}

The CURL block, illustrated in Figure~\ref{fig:DIFAR_pipeline}, is useful in two image enhancement scenarios, both of which we explore in this paper: a) {\em Photo Enhancement (RGB-to-RGB mapping)}: taking an input RGB image and mapping that image to a visually pleasing output RGB image and b) {\em RAW-to-RGB mapping:} modelling the full ISP pipeline, taking RAW sensor input and producing an RGB output image. 

Photo enhancement typically relates to brightness and colour adjustment. The state-of-the-art DeepUPE~\cite{wang19} approach learns a luminance adjustment matrix that improves the exposure of an image. Hu~\etal~\cite{hu2018} design a photo retouching approach (White-Box) using reinforcement learning and GANs to apply a sequence of filters to an image. Deep Photo Enhancer (DPE)~\cite{Chen2018DPE} is a GAN-based architecture with a U-Net-style generator for RGB image enhancement that produces state-of-the-art results on the popular MIT-Adobe5K benchmark dataset. Recently~\cite{GuoCVPR2020} demonstrated an image enhancement method that does not require reference training images, and~\cite{BiancoCVPRW2019} presented a content-preserving tone adjustment method. Arguably, the highest quality in digital photography is achieved with DSLR which, due to the large aperture and lenses, can produce higher quality photographs than 
a smartphone camera. Related work enhances smartphone images by learning a mapping between smartphone images and DSLR~\cite{IgnatovICCV2017,zoom_CVPR2019} photographs using deep learning. Our proposed CURL block differs from these methods in its exploitation of multiple colour spaces and the loss function which drives curve learning jointly in each colour space, outperforming~\cite{wang19,Chen2018DPE,IgnatovICCV2017} on the RGB-to-RGB task.

Further combination of multiple ISP tasks (\eg denoising, demosaicing, photo enhancement \etc) ultimately leads to modeling the entire ISP RAW-to-RGB transformation. Currently the literature is limited for deep learning methods that 
replace the full pipeline. The most relevant related work is 
DeepISP~\cite{schwartz19}, which incorporates a low level network that performs local adjustment of the image including joint denoising and demosaicing, and a high level network that performs global image enhancement. Other recent work includes the Self-Guided Network (SGN)~\cite{Gu_2019_ICCV} which relies extensively on pixel shuffling~\cite{Ledig17} operations, and CameraNet~\cite{cameraNet_2019} that decomposes the problem into two subproblems of restoration and enhancement. Another related work is~\cite{Chen2018} which also learns a RAW-to-RGB mapping, however for extremely dark images using a simple U-Net-style architecture~\cite{Ronneberger15}. Nam and Kim~\cite{Nam2017ModellingTS} present a deep model that mimicks the RAW-to-RGB ISP for different cameras. Despite this progress, it is an open question how far image quality can advance using deep RAW-to-RGB networks. Similar to~\cite{Chen2018}, our proposed pixel-level block has an encoder/decoder structure, however we replace standard skip connections with novel multi-scale contextual awareness (MSCA) connections, providing enriched features from the encoder to the decoder that boost performance. CURL includes global image transformations to adjust colours and brightness, 
similar to~\cite{schwartz19}, however, we introduce neural curve layers, which more expressively adjust image properties in a controlled but human-interpretable way.  Unlike~\cite{Chen2018} or~\cite{schwartz19}, the CURL block 
uses multiple colour spaces, producing results that outperform~\cite{schwartz19} and~\cite{Chen2018}
on the RAW-to-RGB mapping.

\section{TED $+$ CURL for Image Enhancement}\label{sec:methods}

CURL receives and processes convolutional features from a backbone network (Figure~\ref{fig:DIFAR_pipeline}). The backbone network, dubbed \textbf{T}ransformed \textbf{E}ncoder-\textbf{D}ecoder (TED), is a multi-scale encoder-decoder neural network (Section \ref{sec:lowlevel}) for local pixel processing. The output of TED is passed to the CURL block  (Section~\ref{sec:highlevel}), employing  novel \emph{neural curve layers} that globally adjust image properties. We argue that, as in previous work~\cite{schwartz19,Chen2018DPE}, effective image enhancement requires both local (Section~\ref{sec:lowlevel}) and global (Section~\ref{sec:highlevel}) adjustment. In Section~\ref{sec:lowlevel} we describe TED,  used for local pixel processing, while in Section~\ref{sec:highlevel} we describe CURL. The combination of both architectures is referred to as TED$+$CURL.

\subsection{Transformed Encoder-Decoder (TED) for Local Image Adjustment}
\label{sec:lowlevel}

\begin{figure}[t!]
\begin{center}
\includegraphics[scale = 0.185]{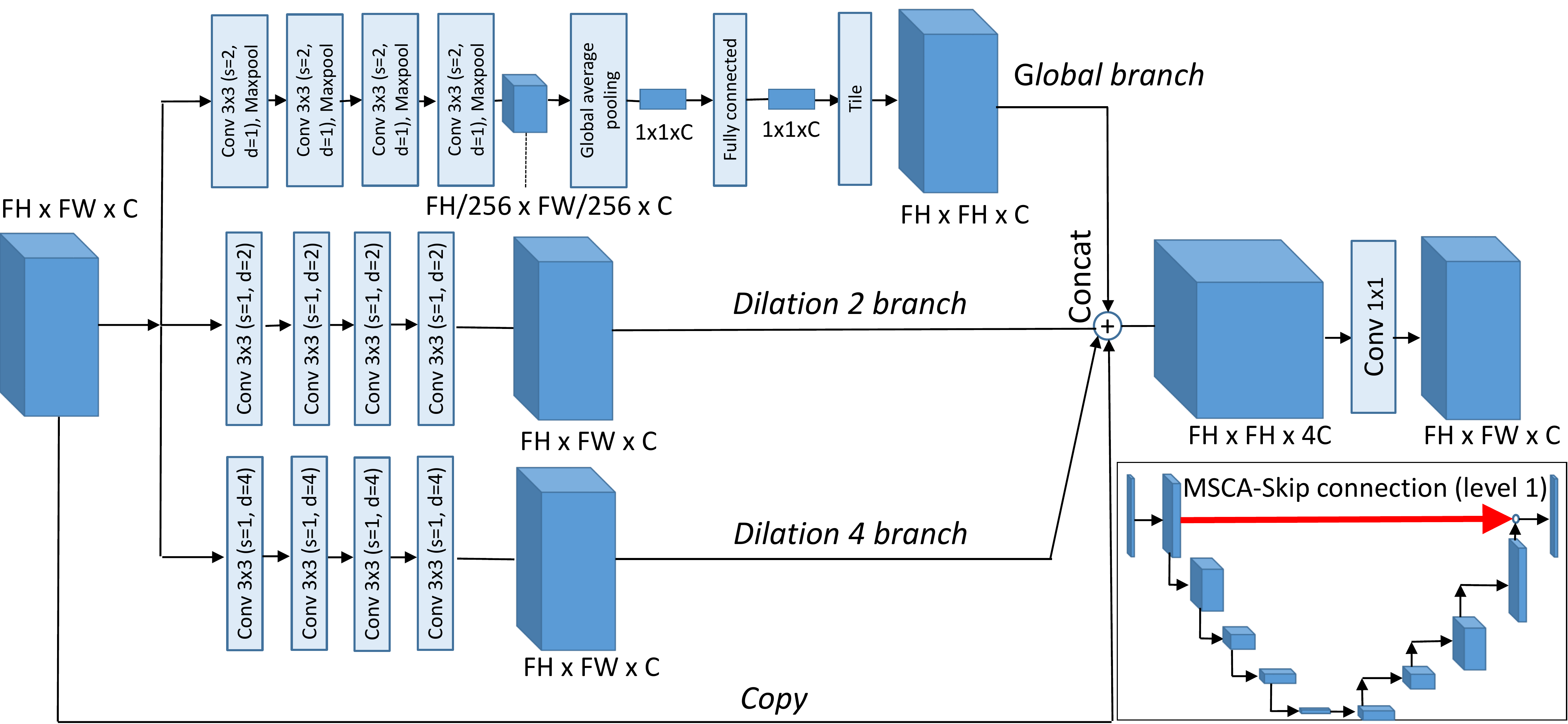}
\end{center}
\caption{\textbf{Transformed encoder/decoder (TED) backbone} fusing multiple levels of image context to deliver more contextually relevant features for the expanding path (Section~\ref{sec:lowlevel}). \textbf{Main image:} The multi-scale contextual awareness skip connection. \textbf{Bottom right:} the connection is placed at the first level skip connection. All other standard U-Net skip connections are removed.}
\label{fig:fusion_block}
\end{figure} 

\begin{figure}[t!]
\begin{center}
\includegraphics[width=84mm,height=50mm]{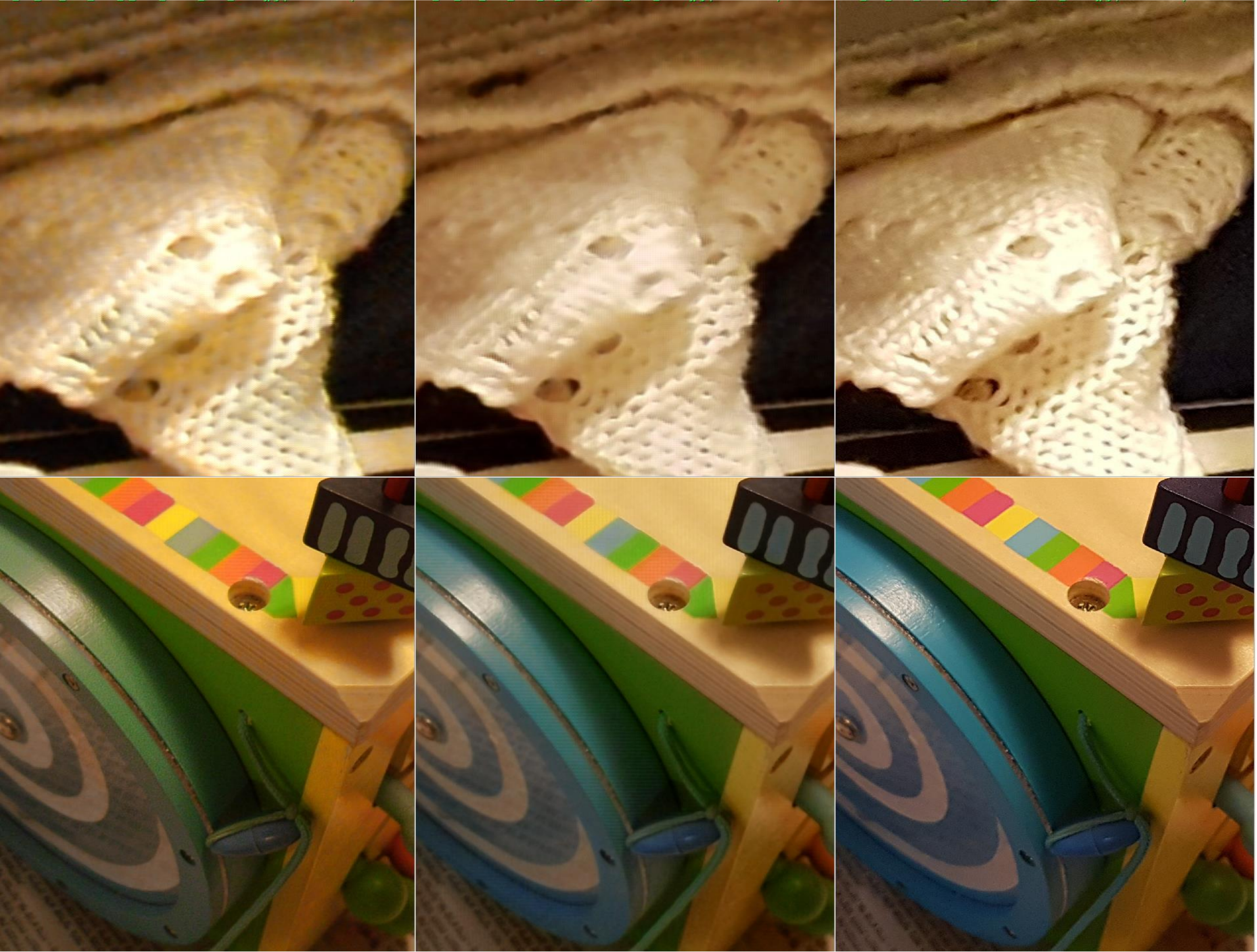}
\end{center}
\caption{\textbf{Left:} U-Net. \textbf{Middle:} TED. \textbf{Right:} Groundtruth.  TED improves colour reproduction.}
\label{fig:color_bleeding}
\end{figure}

Despite its popularity, in our experimental evaluation (Table~\ref{tab:resultsMCSA2}), we find that the standard U-Net backbone~\cite{Ronneberger15,Chen2018DPE,Chen2018} is not entirely suitable for the image translation task, as it is important to consider global and mid-level context when making local pixel adjustments to reduce spatial inconsistencies in the predicted image. To address this, we redesign the encoder / decoder backbone with a skip connection (Figure~\ref{fig:fusion_block}) that we call \emph{multi-scale contextual awareness} (MSCA) connection. Our \textbf{T}ransformed \textbf{E}ncoder-\textbf{D}ecoder is dubbed \textbf{TED}. The MSCA-connection fuses multiple different contextual image features, combining global and mid-level features to enable cross-talk between image content at different scales. The MSCA-connection uses convolutional layers with dilation rate 2 and 4 to gain a larger receptive field and capture mid-level image context from the input. These mid-level feature maps are at the same spatial resolution as the input tensor to the block. Global image features are extracted using a series of convolutional layers with a stride 2, followed by a Leaky ReLU activation and then a max pooling operation. These layers are then followed by global average pooling and a fully connected layer, which outputs a fixed dimensional feature vector, tiled across the height and width dimensions of the input. The resulting feature maps are concatenated to the input and fused by a $1$x$1$ convolution to produce a tensor with much fewer channels. This merges the \emph{local, mid-level and global} contextual information for concatenation to the feature maps of the upsampling path at that particular level. We find that a MSCA-skip connection at level one of the encoder/decoder provides a good trade-off between parameter complexity and image quality, with all other skip connections removed (Figure~\ref{fig:fusion_block}). Multi-scale neural blocks are not new, having been suggested in the literature before~\cite{gharbi2017deep,Chen2018,Marnerides18}. However, to the best of our knowledge, we are the first to show that \emph{endowing a skip connection} with a multi-scale processing block can dramatically reduce the number of parameters required while increasing the image quality (Table~\ref{tab:resultsMCSA2}, Figure~\ref{fig:color_bleeding}) for RAW-to-RGB and RGB-to-RGB mapping. 

\begin{table}[t]
\centering
\caption{Backbone network (TED) with MSCA-skip connection, improves image quality with fewer parameters (Samsung S7 dataset).}
\adjustbox{max width=\textwidth}{%
\begin{tabularx}{\linewidth}{ l | c c c }
\multicolumn{4}{c}{} \\
\textbf{Architecture} & \textbf{PSNR} & \textbf{SSIM} & \textbf{\# Parameters} \\
\hline
TED (MSCA-skip, level 1) & ${\textbf{26.56}}$     &${0.781}$    &    \textbf{1.3 M} \\ 
TED (MSCA-skip, all levels) & ${26.39}$     &${\textbf{0.793}}$    &    {3.3} M \\ 
\hline
\hline
U-Net          & ${25.78}$ & ${0.771}$ & 1.4 M \\ 
U-Net (large)  & ${25.37}$ & ${0.788}$ & 5.1 M \\  
\hline
\end{tabularx}
}
\label{tab:resultsMCSA2}
\end{table}

Different to previous work, we evaluate our approach on \emph{both} RAW-to-RGB and photo enhancement (RGB-to-RGB). The backbone encoder/decoder differs depending on the type of input data (RAW or RGB). Both architecture diagrams are presented in the supplementary material. For RAW-to-RGB experiments we modify the encoder architecture to facilitate the processing of RAW images represented by a colour filter array (CFA). The CFA tensor $H{\times}W{\times}1$ is shuffled into a packed form using a pixel shuffle layer~\cite{Ledig17} to give a $(H/r)(W/r){\times}r^{2}$ tensor, where we set $r{=}2$ in this work. This tensor is fed into the downsampling path of the backbone network. The output from the expanding path is a set of feature maps of half the width and height of the full RGB image. These feature maps are added to the input packed image (replicated $4{\times}$ on channel dimension) through a long skip connection. A pixel shuffle upsampling operation~\cite{Ledig17} produces a result with output shape $H{\times}W{\times}C$, where $C$ is the number of feature maps, and is passed to the retouching block for further processing. Our RGB-to-RGB network is broadly similar to the RAW-to-RGB network, however the pixel shuffling operations are removed. The long skip connection remains, but in this case the 3 channel input image is added to the output, rather than a shuffled version. 

\begin{figure*}[t!]
\begin{center}
\includegraphics[scale = 0.4]{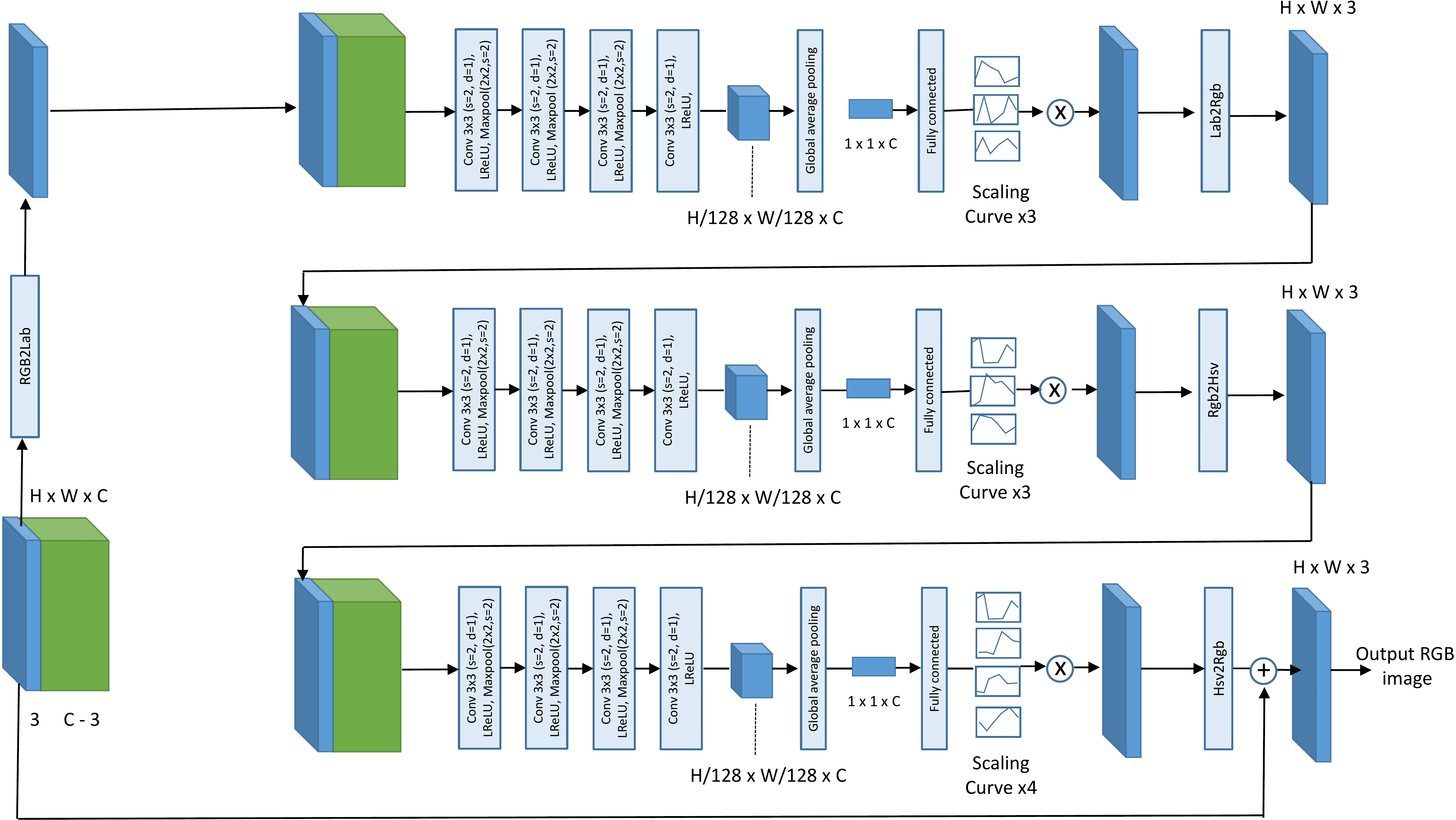}
\end{center}
\caption{\textbf{CURL block} refines the image from the backbone network to adjust luminance, colour and saturation. The image has luminance adjusted first, color (RGB) second, and saturation third using three piecewise linear curves. The output of this block is the final RGB image. $C$ is number of feature maps, $s$ stride, $d$ dilation rate, $H$ image height, $W$ image width.}
\label{fig:neural_block}
\end{figure*}

\subsection{CURL: Neural Curve Layers for Global Image Adjustment}\label{sec:highlevel}

This section describes CURL, a pluggable global image colour and luminance retouching block which forms the central contribution of this paper. The first three channels of the $H{\times}W{\times}C$ tensor from the pixel-level block are treated as the image to be globally adjusted, and the remaining channels serve as feature maps that are used as input to each \emph{neural curve layer}. Our proposed neural curve layer block is shown in Figure~\ref{fig:neural_block} and consists of, for each of the three colour spaces, a global feature extraction block followed by a fully connected layer that regresses the knot points of a piecewise-linear curve (Figure~\ref{fig:DIFAR_pipeline}). The curve adjusts the predicted image ($\hat{I}_{i}\in[0,1]$) by scaling pixels with the formula presented in Equation~\ref{eq:neural_curve}.
\begin{align}
\mathcal{S}(\hat{I}^{jl}_{i}) &= k_{0} + \sum^{M-1}_{m=0}(k_{m+1}-k_{m})\delta(M\hat{I}^{jl}_{i}-m) \label{eq:neural_curve}, \\      &\text{where}\hspace{1em} 
      \delta(x){=}\begin{cases} \nonumber 
      0 & x < 0 \\
      x & 0\leq x\leq 1 \\
      1 & x > 1 \\
        \end{cases}
\end{align}

\begin{figure}[t!]
\begin{center}
\includegraphics[width=0.49\linewidth]{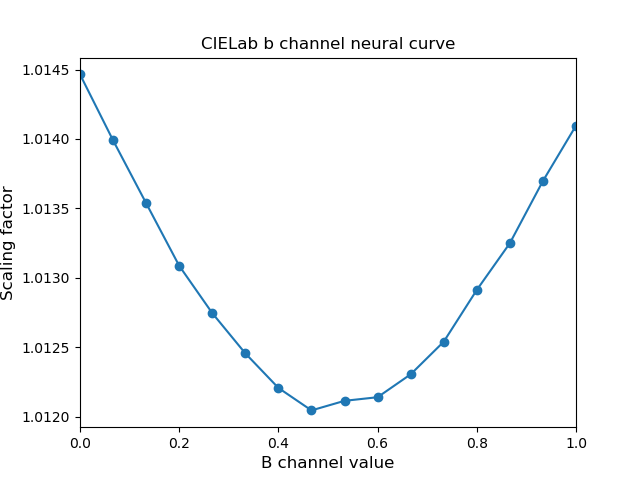}
\includegraphics[width=0.49\linewidth]{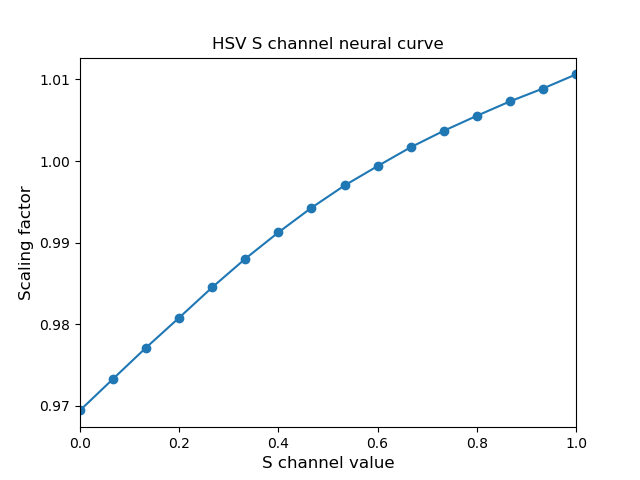}
\end{center}
\caption{Examples of learnt neural global adjustment curves for different colour spaces (CIELab and HSV). Retouching is often a subtle operation, mildly adjusting colour and luminance properties of an image produced by the backbone network. A unique adjustment curve is predicted per image, with the magnitude of the adjustment depending on how much retouching is required for that image.}
\label{fig:neural_curve_examples}
\end{figure}

{\flushleft{where $M$ is the number of predicted knot points, $\hat{I}^{jl}_{i}$ is the $j$-th pixel value in the $l$-th colour channel of the $i$-th image, $k_{m}$ is the value of the knot point $m$. The neural curve outputs scale factors, so to apply the curve is a simple matter of multiplication of a pixel's value with its scale factor indicated by the curve.
Example curves learnt by an instance of our model are shown in Figure~\ref{fig:neural_curve_examples}. CURL 
regresses expressive curves, 
used to \emph{scale} rather than \emph{remap} colors in comparison to existing approaches, e.g.~\cite{GuoCVPR2020,hu2018,BiancoCVPRW2019}.\footnote{Thanks to the scaling, CURL can easily mix representations, for example mapping saturation as a function of hue, whereas previous approaches are restricted to mapping like to like (e.g. red channel to red channel).} The feature extraction block of the neural curve layer (Figure~\ref{fig:neural_block}) accepts a $H{\times}W{\times}C$ feature map, passing it to a group of blocks each consisting of a convolutional layer with $3{\times}3$ kernels of stride $2$, and $2{\times}2$ maxpooling. We place a global average pooling layer and fully connected layer at the end. The neural curves are learnt in several colour spaces (RGB, Lab, HSV).  }}

In Figure~\ref{fig:neural_block}, we arrange the neural curve layers in a particular sequence, adjusting firstly luminance and the a, b chrominance channels (using three curves respectively) in CIELab space. Afterwards, we adjust the red, green, blue channels (using three curves respectively) in RGB space. Lastly hue is scaled based on hue, saturation based on saturation, saturation based on hue, and value based on value (using four curves respectively) in HSV space. This ordering of the colour spaces is found to be optimal based on a sweep on a validation dataset\footnote{An experimental study on the ordering of the colour spaces is presented in Section~\ref{sec:experiments}.}. The input to each neural curve layer consists of the concatenation of a $H{\times}W{\times}3$ image converted to the given colour space, and a $H{\times}W{\times}C'$ subset of the features from the pixel-level block ($C'{=}C{-}3$). For the luminance neural curve layer, the fully connected layer regresses the parameters of the L, a and b channel scaling curves. The scaling curves scale the pixel values in the L, a, b channels using Equation~\ref{eq:neural_curve}. The adjusted CIELab image is then converted back to RGB. This $H{\times}W{ \times}3$ RGB image is concatenated with the pixel-level block $H{\times}W{\times}C'$ feature map and fed into the second neural curve layer that learns three more curves, one for each channel of the RGB image. These curves are applied to the R, G, B channels to adjust the colours. Lastly, the $H{\times}W{\times}3$ RGB image is converted to HSV space. HSV space separates the hue, saturation and value properties of an image. The HSV image is concatenated with the $H{\times}W{\times}C'$ feature map and used by the final curve layer to predict the four HSV space adjustment curves. The four HSV curves are for saturation-to-saturation mapping, hue-to-hue, value-to-value and hue-to-saturation mapping. The hue-to-saturation curve is particularly interesting as it permits the precise adjustment of the saturation of individual hues. These curves are applied to the HSV image and the HSV image is converted back to RGB space via a differentiable HSV to RGB conversion. A long skip connection links input and output. The differentiable colour spaces transformations required in the CURL block are readily implementable using~\eg the Kornia library~\cite{Riba_2020_WACV}.

\subsection{The CURL Loss Function}\label{sec:loss}

The CURL loss function consists of three colour space-specific terms which seek to optimise different aspects of the predicted image, including the hue, saturation, luminance and chrominance. The loss is designed to control each of the colour-space specific transformations in CURL. The loss is minimised over a set of $N$ image pairs $\{(I_{i},\hat{I}_{i})\}^{N}_{i=1}$, where $I_{i}$ is the reference image and $\hat{I}_{i}$ is the predicted image. The CURL loss is presented in Equation~\ref{eq:difar_loss}
\begin{eqnarray}
	\mathcal{L} = \sum^{N}_{i=1} \mathcal{L}_{hsv}^{i} + \mathcal{L}_{lab}^{i} + \mathcal{L}_{rgb}^{i} + \mathcal{L}_{reg}^{i}
	\label{eq:difar_loss}
\end{eqnarray}
{\flushleft{where $\mathcal{L}_{rgb}$, $\mathcal{L}_{lab}$, $\mathcal{L}_{hsv}$ are the various colour space loss terms, and $\mathcal{L}_{reg}$  is a curve regularization loss.  These terms are defined in more detail below.}}

{\flushleft{\textbf{HSV loss}}}, $\mathcal{L}_{hsv}$: given the hue (angle) $H_{i}\in\left[0,2\pi\right)$, saturation $S_{i}\in\left[0,1\right]$ and value $V_{i}\in\left[0,1\right]$ for image $I_{i}$, we compute $\mathcal{L}_{hsv}^{i}$ in the conical HSV colour space,

\begin{align}
\mathcal{L}_{hsv}^{i} &= \omega_{hsv}(\Vert \hat{S}_{i}\hat{V}_{i}cos(\hat{H}_{i})-{S}_{i}{V}_{i}cos(H_{i})\Vert_{\xstrut 1}  + \nonumber \\ & \Vert\hat{S}_{i}\hat{V}_{i}sin(\hat{H}_{i})-{S}_{i}{V}_{i}sin(H_{i})\Vert_{\xstrut 1})
       \label{eq:hsvloss}
\end{align}
HSV is advantageous as it separates colour into useful components (hue, saturation, intensity). We believe this is one of the first applications of a differentiable HSV transform and loss in a deep network for the purposes of image enhancement. 

\begin{figure*}[!tbp]
\centering
\begin{tabular}{c@{}c@{}c@{}c@{}c@{}}
      \scalebox{0.70}{$\mathcal{L}_{lab} {+} \mathcal{L}_{reg}$} & 
      \scalebox{0.70}{$\mathcal{L}_{lab}  {+} \mathcal{L}_{hsv} {+} \mathcal{L}_{reg} $}  &
      \scalebox{0.67}{$\mathcal{L}_{lab} {+} \mathcal{L}_{hsv} {+}  \mathcal{L}_{rgb,no{-}cos} {+} \mathcal{L}_{reg}$} & 
      \scalebox{0.70}{All terms} &
      \scalebox{0.70}{Groundtruth} \\
     \includegraphics[width=0.198\linewidth]{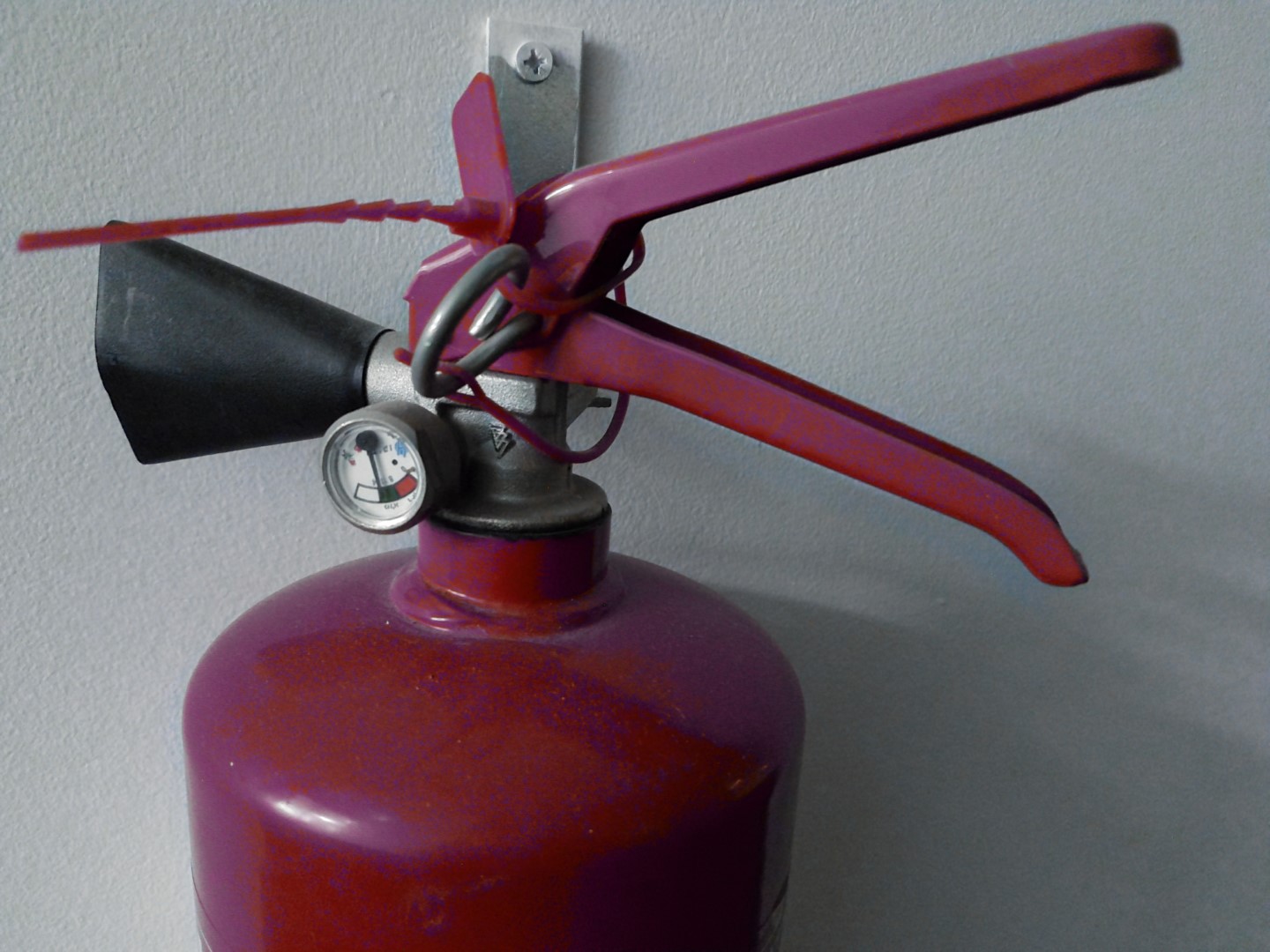} & 
     \includegraphics[width=0.198\linewidth]{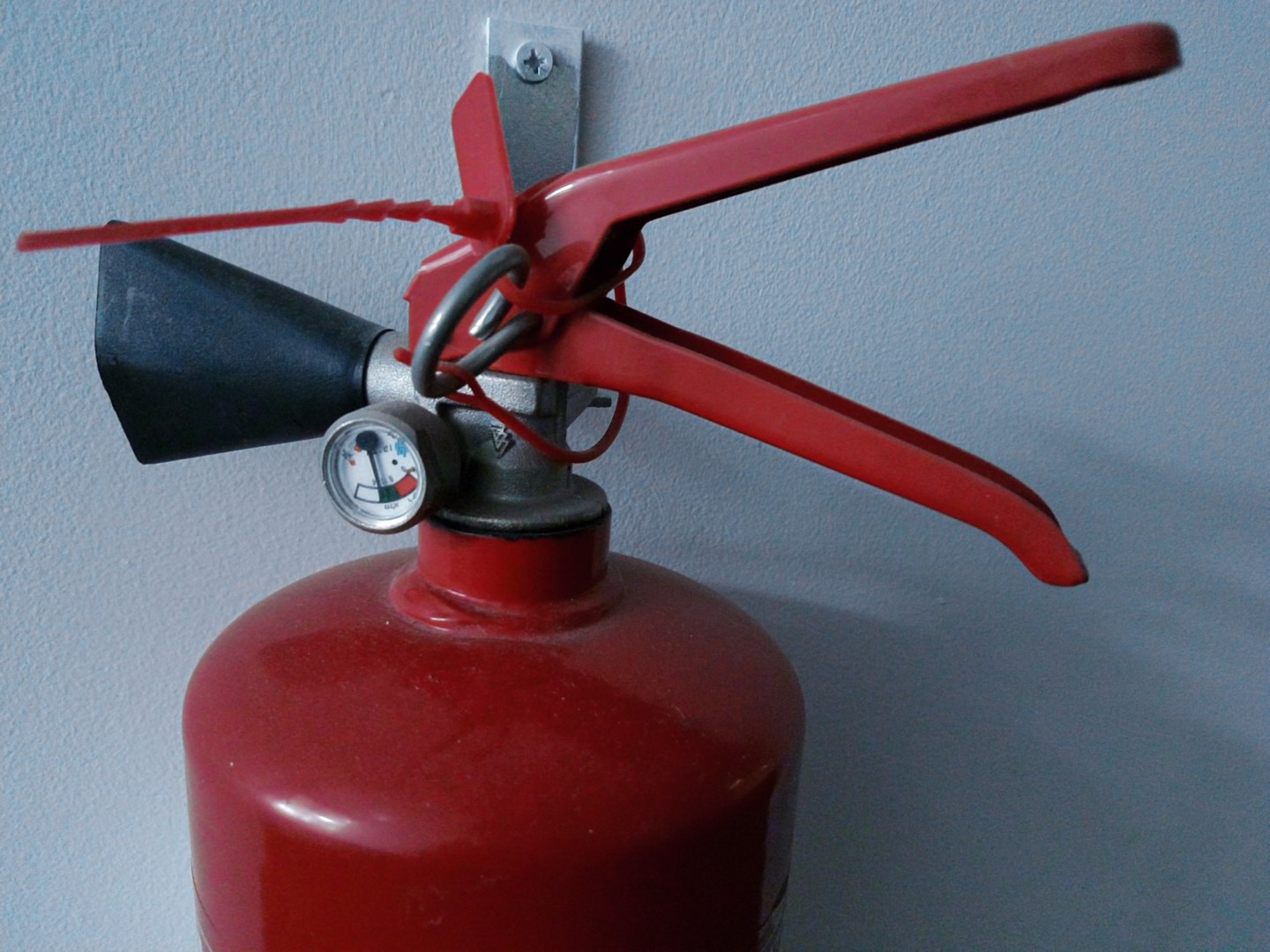}&  
     \includegraphics[width=0.198\linewidth]{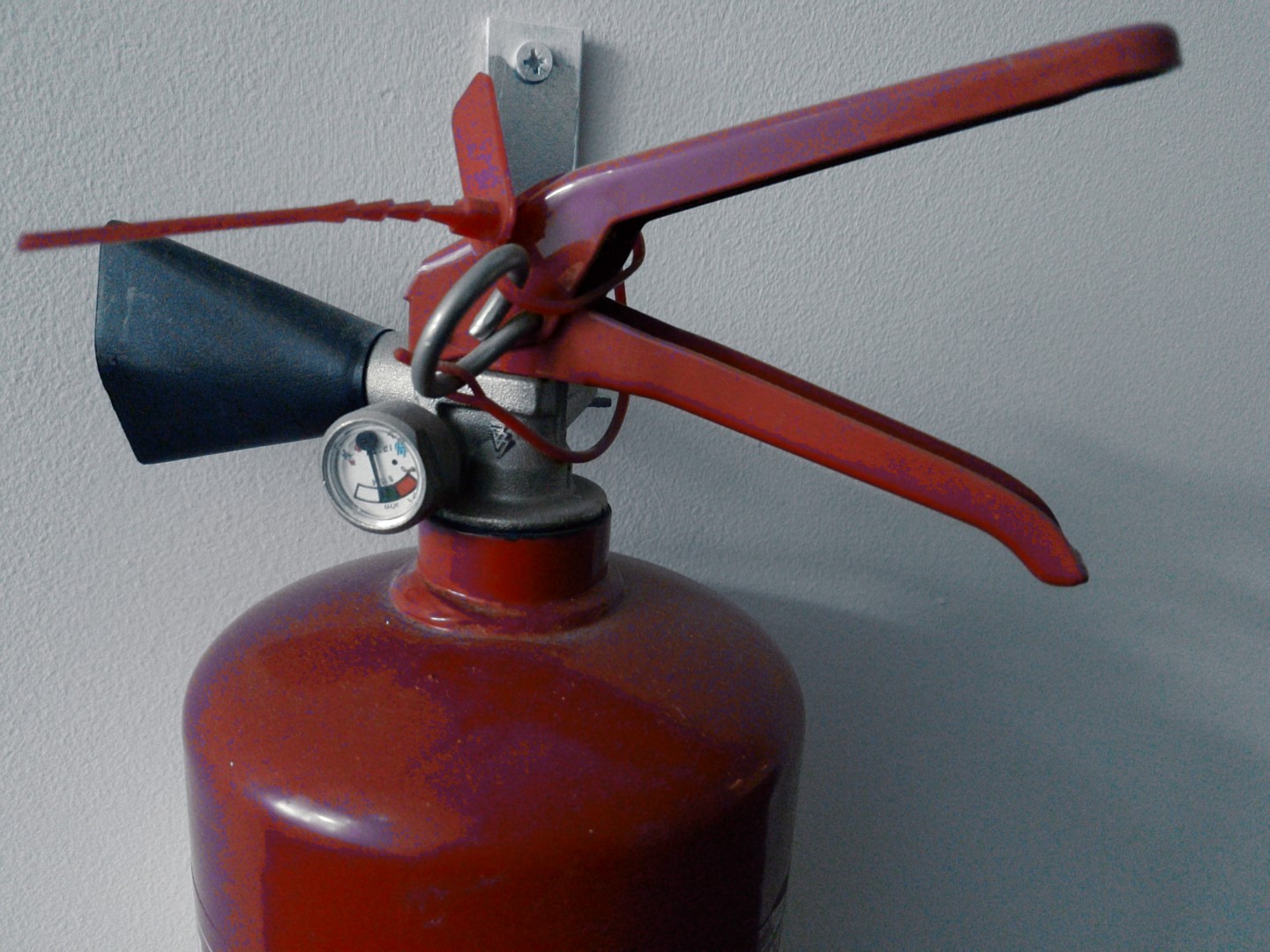} &
     \includegraphics[width=0.198\linewidth]{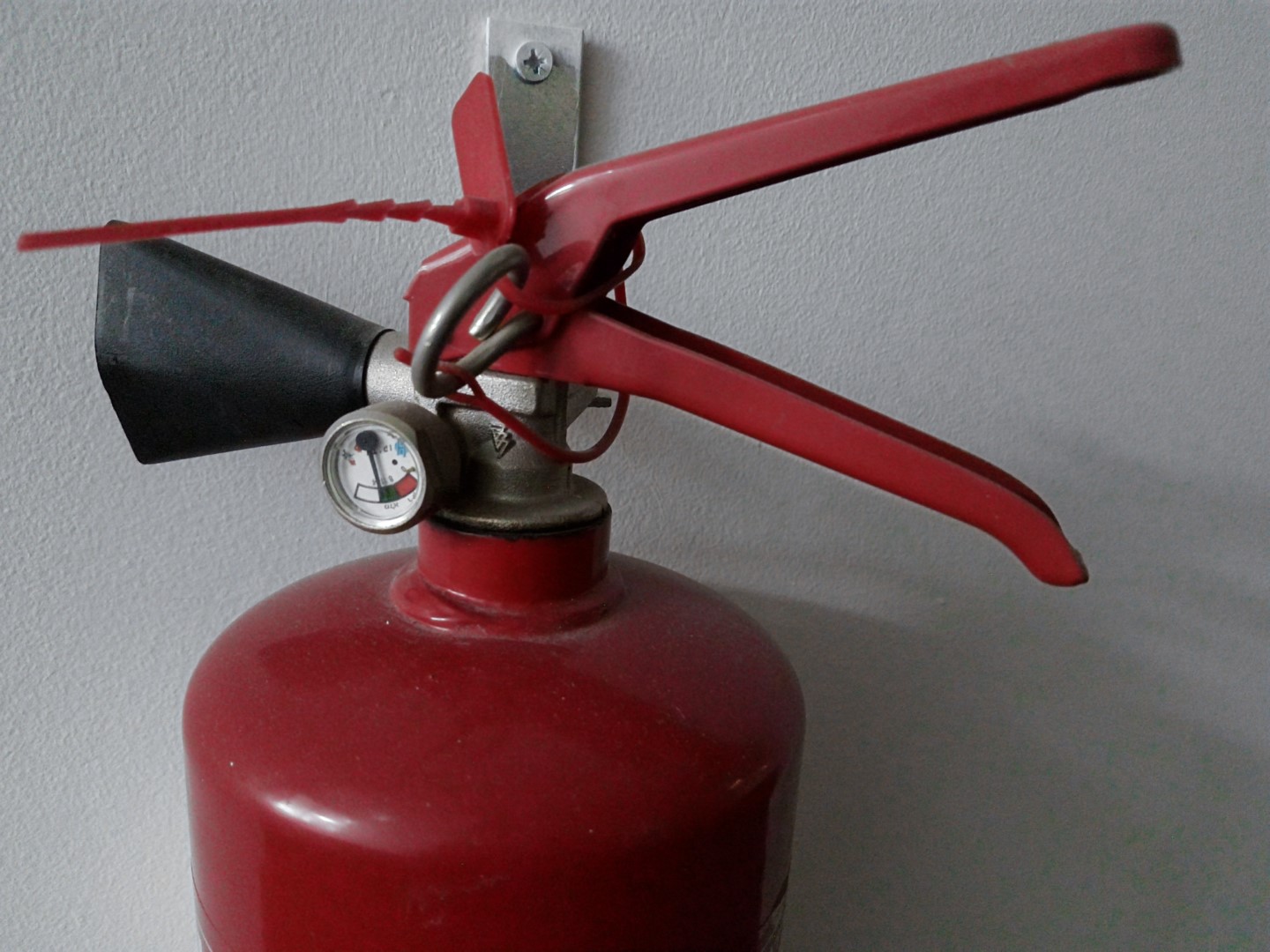} &
    \includegraphics[width=0.198\linewidth]{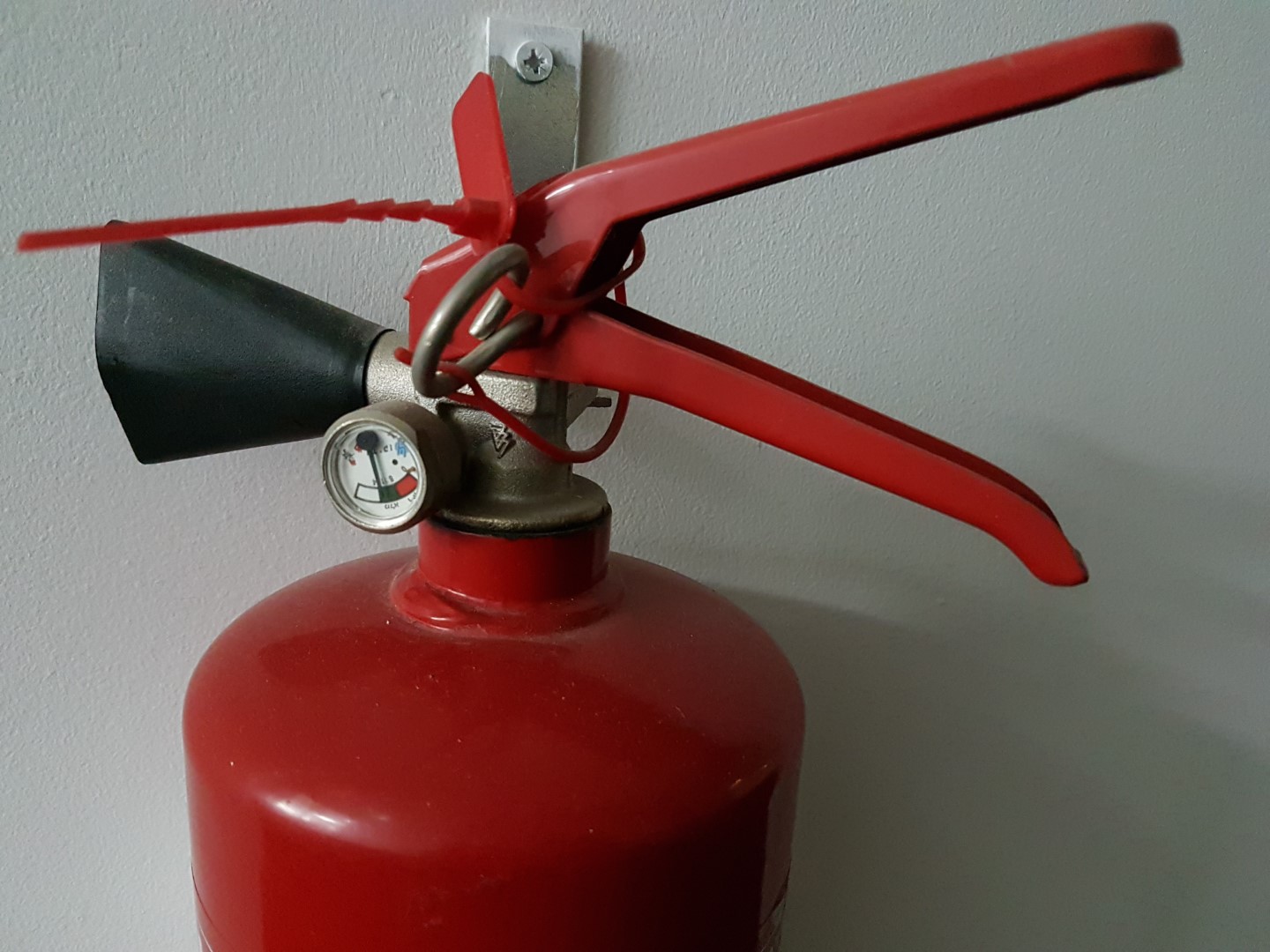} \\
\end{tabular}
    \caption{Qualitative effect of different combinations of terms in the CURL loss function on image quality. $\mathcal{L}_{rgb,no{-}cos}$ is the RGB loss without the cosine distance term. All terms are most effective, with obvious artefacts and colour distortions removed. See Section~\ref{sec:loss} for more detail.}
    \label{fig:loss_ablation}
\end{figure*}

{\flushleft{\textbf{CIELab loss}}}, $\mathcal{L}_{lab}$: we compute the $L_{1}$ distance between the Lab values of the groundtruth and predicted images (Equation \ref{eq:labloss}). In particular, the multi-scale structural similarity (MS-SSIM)~\cite{wang03} between the luminance channels of the ground truth and predicted images enforces a reproduction of the contrast, luminance and structure of the target image~\cite{schwartz19}. 
\begin{align}
\mathcal{L}_{lab}^{i} &= \omega_{lab}\Vert Lab(\hat{I}_{i})-Lab(I_{i})\Vert_{\xstrut 1}  + \nonumber \\ & \omega_{ms{-}ssim}\text{MS-SSIM}(L(\hat{{I}}_{i}),L(I_{i}))
       \label{eq:labloss}
\end{align}

{\flushleft{where $Lab(.)$ is a function that returns the CIELab Lab channels corresponding the RGB channels of the input image and $L(.)$ is a function that returns the L channel of the image in CIELab colour space.}}

{\flushleft{\textbf{RGB loss}}}, $\mathcal{L}_{rgb}$: this term consists of $L_{1}$ distance on RGB pixels between the predicted and groundtruth images and a cosine distance between RGB pixel vectors (where ${I}^j_i$ is three-element vector representing the RGB components of pixel $j$ in the $i$-th image)
\begin{equation}
       \mathcal{L}_{rgb}^{i} = \omega_{rgb}\Vert\hat{I}_{i}-I_{i}\Vert_{\xstrut 1} + \omega_{cosine}(1-\frac{1}{HW}\sum^{HW}_{j=1}(\frac{\hat{I}^{j}_{i}.I^{j}_{i}}{\Vert\hat{I}^{j}_{i}\Vert_{\xstrut 2}\Vert I^{j}_{i}\Vert_{\xstrut 2}})) 
      \label{eq:rgbloss}
\end{equation}
 

{\flushleft{\textbf{Curve regularization loss}}}, $\mathcal{L}_{reg}$: to mitigate overfitting we regularize by penalizing the curvature:

\begin{equation}
       \mathcal{L}_{reg}^{i} = \omega_{reg}\sum^{R-1}_{r=0}\sum^{M-3}_{m=0}(\Delta{k}^{r}_{m+1}-\Delta{k}^{r}_{m})^{2}
      \label{eq:regloss}
\end{equation}
 
{\flushleft{where $\Delta{k}_{m}$ is the gradient of the line segment defined by knot points $k_{m},k_{m+1}$, $M$ is the number of knot points, and $R$ denotes the number of curves.}}

{\flushleft{\textbf{Loss term weights:}}} each loss term has an associated weight hyperparameter: $\omega_{rgb}$, $\omega_{hsv}$, $\omega_{lab}, \omega_{ms{-}ssim},\omega_{reg}$,$\omega_{cosine}$. We empirically find that only the $\omega_{mssim}$ term is sensitive to the particular dataset. The supplementary material presents details on our training configuration. Our experimental evaluation (Section~\ref{sec:experiments}) presents evidence towards the necessity and contribution of each term, in relation to output image quality. 

\section{Experimental Results}\label{sec:experiments}

{\flushleft{\textbf{Datasets:}\quad We evaluate CURL on three publicly available datasets: \textbf{(i) Samsung S7~\cite{schwartz19}} consists of 110, 12M pixel images of short-to-medium exposure RAW, RGB image pairs and medium-to-medium exposure RAW, RGB pairs. Following~\cite{schwartz19} we divide the dataset into 90 images for training, 10 for validation and 10 for testing. \textbf{(ii) MIT-Adobe5k-DPE~\cite{Chen2018DPE}} contains 5,000 images taken on DSLR cameras and subsequently adjusted by an artist (Artist C). We follow the dataset pre-processing procedure of DeepPhotoEnhancer (DPE)~\cite{Chen2018DPE}. The training dataset consists of 2,250 RGB image pairs. The RAW input images are processed into RGB by Lightroom. The groundtruth RGB images are generated by applying the adjustments of Artist C to the input. The dataset is divided into 2,250 training images and 500 test images. We randomly sample 500 validation images from the 2,250 training images. The images are re-sized to have a long-edge of 500 pixels. \textbf{(iii) MIT-Adobe5k-UPE~\cite{wang19}} MIT-Adobe5k dataset pre-processed as described in~\cite{wang19} (DeepUPE). The training dataset consists of 4,000 images of RGB image input, groundtruth pairs. The groundtruth are from Artist C. The dataset is divided into 4,500 training images and 500 test images. We randomly sample 500 images to serve as our validation dataset. The images are not re-sized and range from 6M pixel to 25M pixel resolution. \textbf{Evaluation Metrics:}~Our image quality metrics are PSNR, SSIM and the perceptual quality aware metric LPIPS~\cite{zhang18}. Hyperparameters were tuned on the held-out validation portion of each benchmark dataset. See supplementary material. \textbf{Backbone:}~For the ablation studies in Tables~\ref{sec:ablation} we use the parameter heavier TED backbone network with an MSCA-skip connection at every level. This variant mostly leads to the highest image quality. For the comparison to the state-of-the-art in Section~\ref{sec:sota} we use the parameter-light variant of TED with one MSCA-skip connection at the first level.} }

\subsection{CURL Ablation Studies}\label{sec:ablation}

\begin{figure*}[!t]
\centering
\begin{tabular}{c@{}c@{}c@{}c@{}c@{}}
      \scalebox{0.85}{HSV  (21.99 dB)} & 
      \scalebox{0.85}{RGB  (22.93 dB)}  &
      \scalebox{0.85}{LAB  (24.76 dB)} & 
      \scalebox{0.85}{All  (25.86 dB)} &
      \scalebox{0.85}{Groundtruth} \\
          \includegraphics[width=0.198\linewidth]{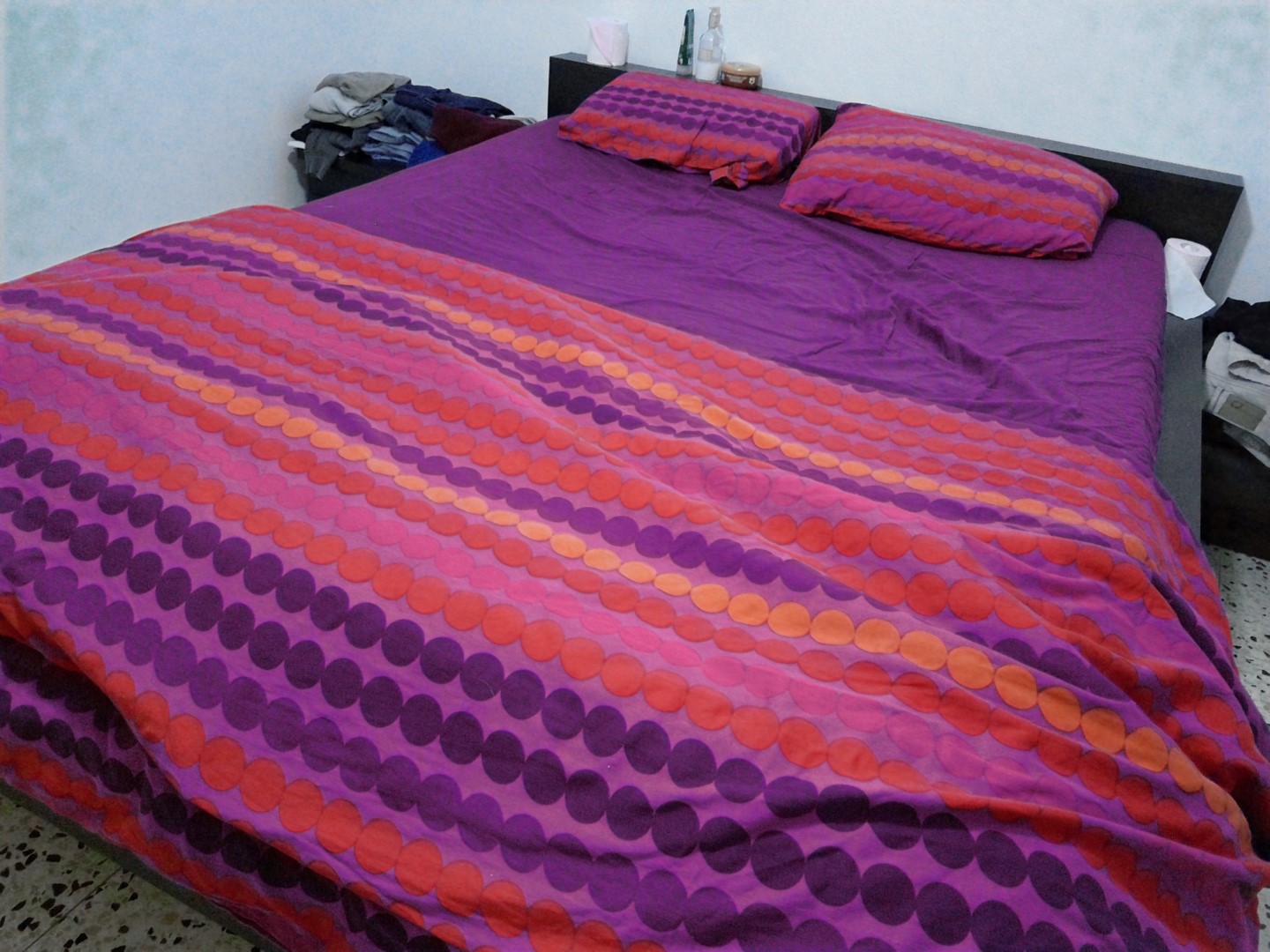} & 
     \includegraphics[width=0.198\linewidth]{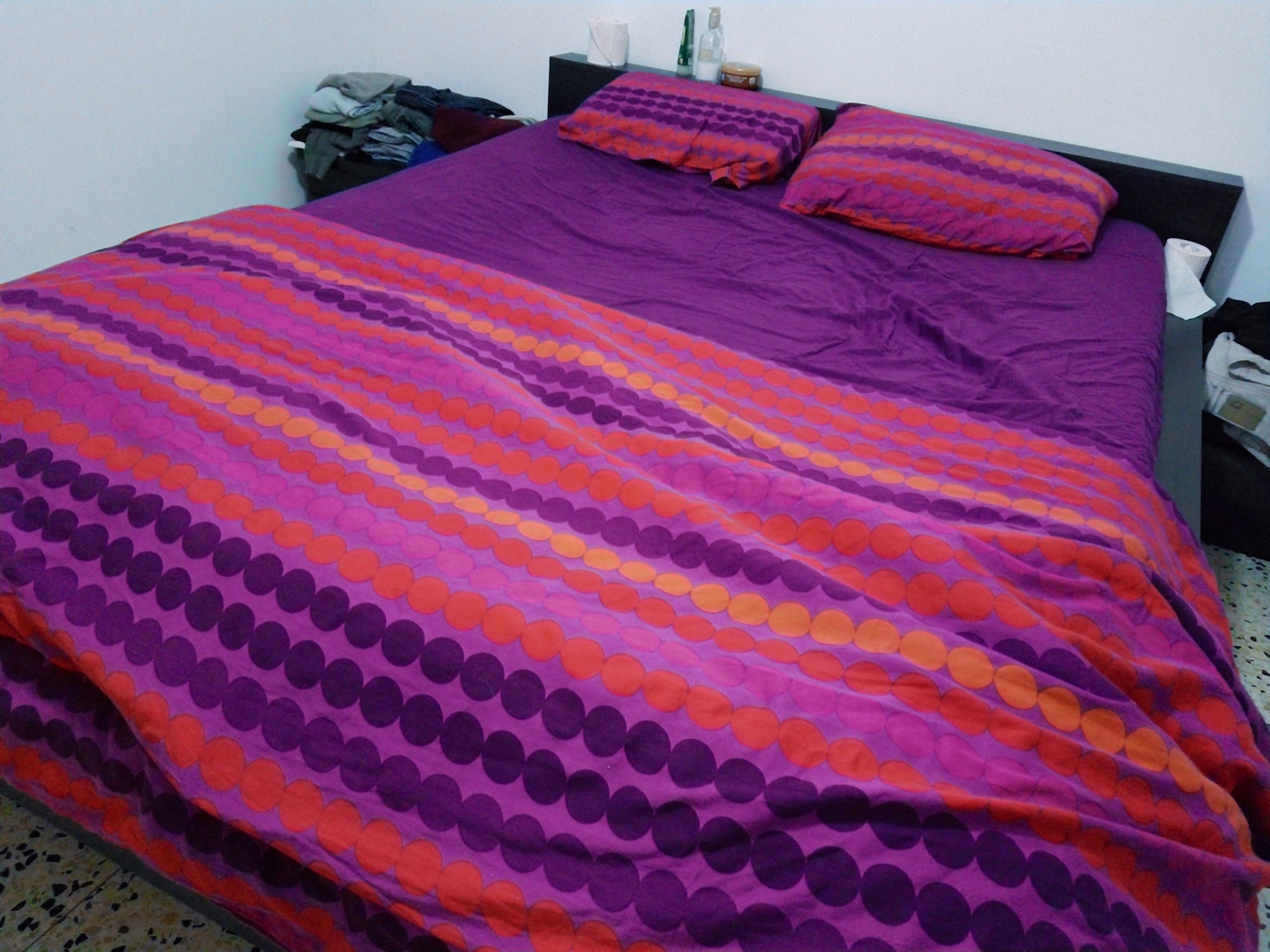} &
     \includegraphics[width=0.198\linewidth]{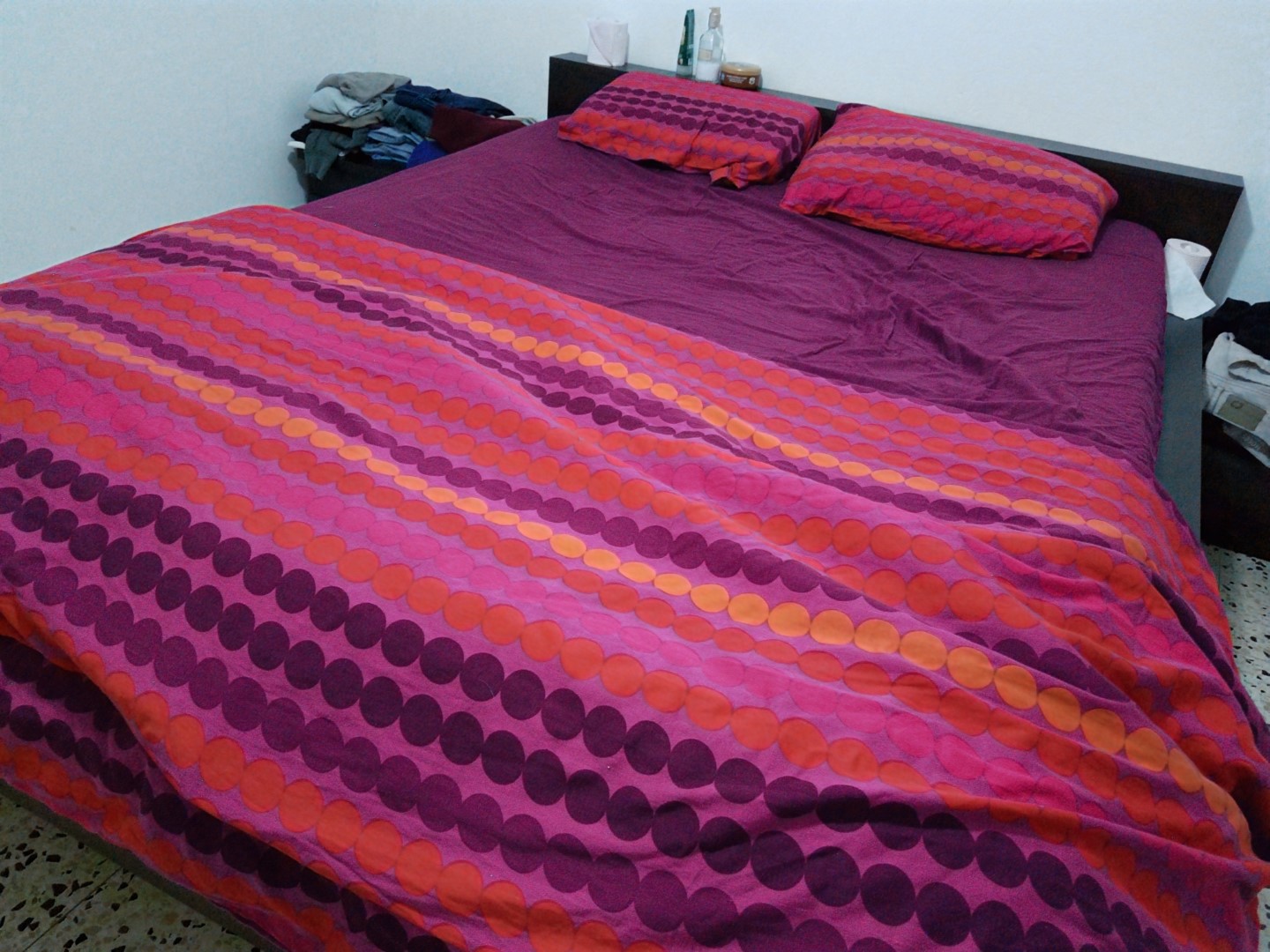} &
     \includegraphics[width=0.198\linewidth]{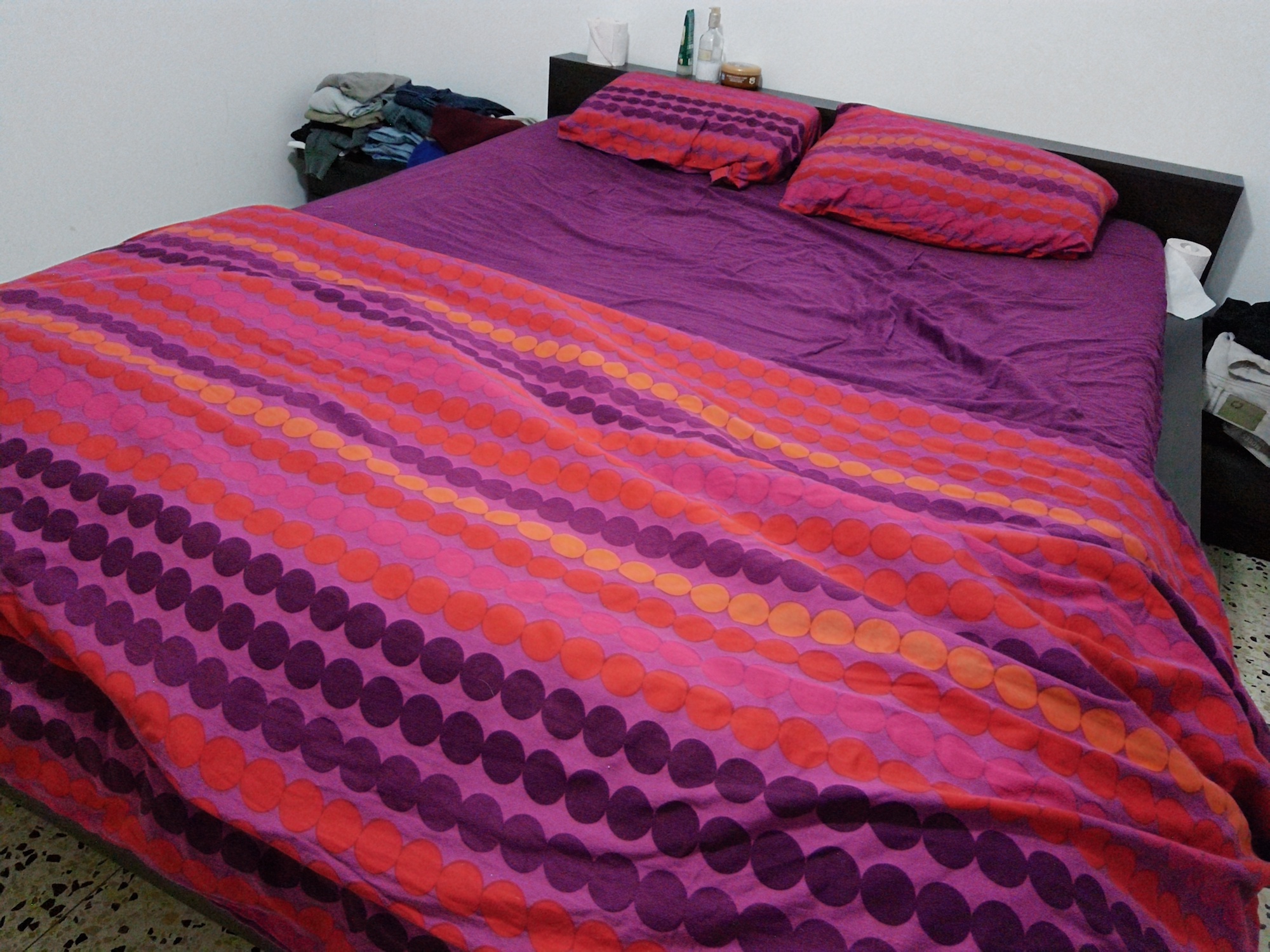} &
    \includegraphics[width=0.198\linewidth]{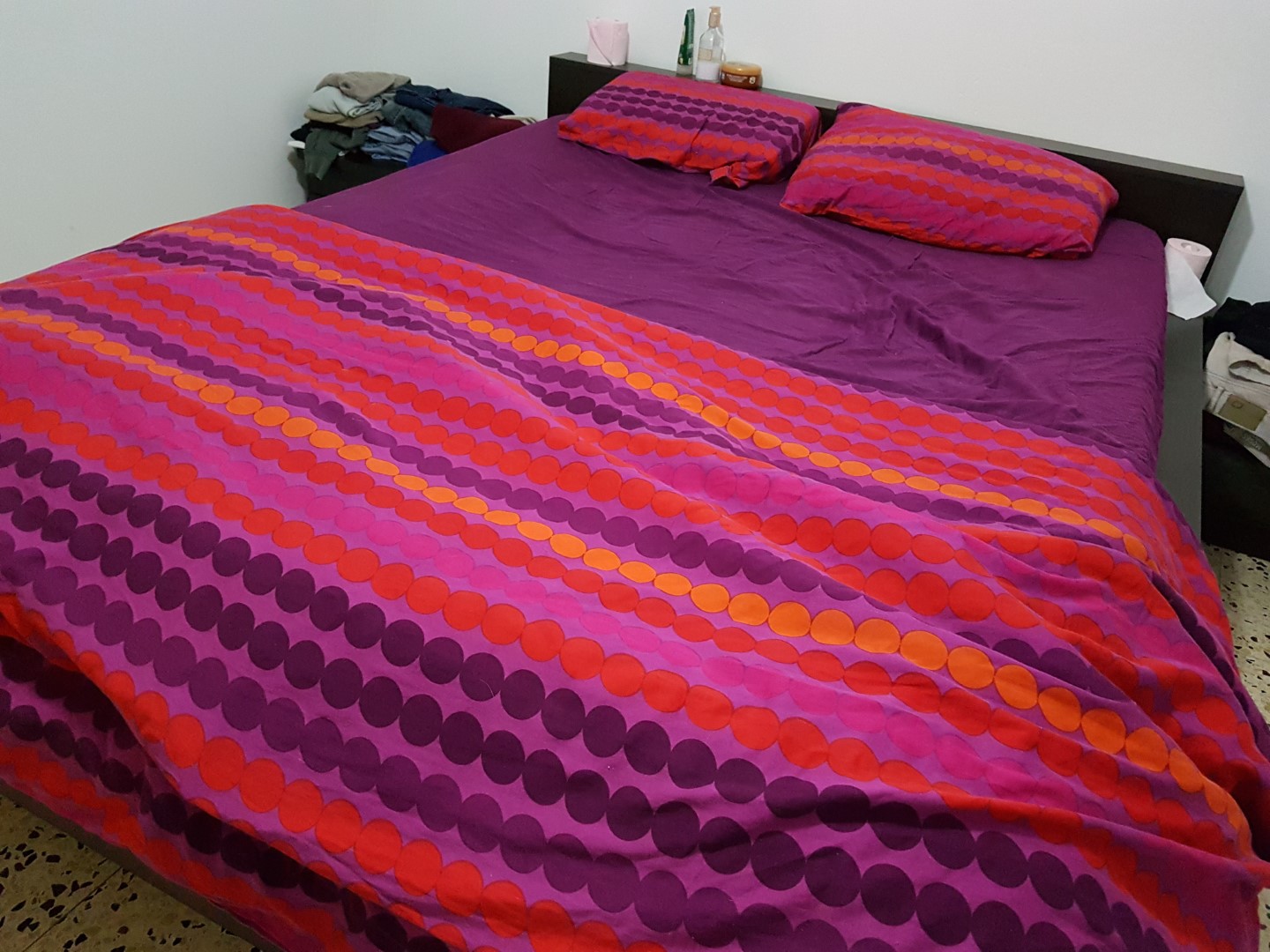} \\
\end{tabular}
    \caption{Qualitative effect of \textbf{different formulations of the CURL block}. We keep only one colour space (either CIELab, RGB, HSV) in the CURL block and compare the output versus keeping all three colour spaces.}
    \label{fig:colour_block_ablation}
\end{figure*}

\begin{table}[t]
\small
\centering
\caption{\small Ablation study on the CURL global image retouching layer.  All colour spaces are necessary in CURL for best image quality.}
\adjustbox{max width=\textwidth}{%
\begin{tabularx}{\linewidth}{ l | X X X } 
\textbf{\small{Architecture}} & \textbf{PSNR}$\uparrow$ & \textbf{SSIM}$\uparrow$ & \textbf{LPIPS}$\downarrow$  \\
\hline
TED$+$CURL (RGB only)         & 26.74          & 0.790          &  0.340 \\ 
TED$+$CURL (HSV only)         & 25.88          & 0.780          & \textbf{0.308} \\
TED$+$CURL (CIELab only)      & 26.98          & 0.788          & 0.323 \\ 
TED$+$CURL (HSV, RGB, CIELab) & \textbf{27.09} & \textbf{0.793} & 0.312 \\ 
\hline
\end{tabularx}
}
\label{tab:color_blocks_ablation}
\end{table}

\begin{table}[t]
\centering
\caption{\small Results for all permutations of the colour spaces in our CURL block.}
\adjustbox{max width=\textwidth}{%
\begin{tabularx}{\linewidth}{ l | X X X X } 
\textbf{Ordering} & \textbf{PSNR (test)}$\uparrow$ & \textbf{SSIM (test)}$\uparrow$ &  \textbf{PSNR (valid)}$\uparrow$  & \textbf{PSNR (valid)}$\uparrow$ \\
\hline
HSV$\rightarrow$RGB$\rightarrow$LAB  & $26.20$ & ${0.779}$  & $26.48$ & ${0.765}$ \\ 
RGB$\rightarrow$HSV$\rightarrow$LAB & $26.83$ & ${0.786}$  & $26.23$ & ${0.770}$  \\ 
LAB$\rightarrow$RGB$\rightarrow$HSV & $\textbf{27.09}$ & $\textbf{0.793}$  & $\textbf{26.56}$ & $\textbf{0.771}$  \\ 
LAB$\rightarrow$HSV$\rightarrow$RGB & $26.37$ & ${0.784}$  & $26.44$ & ${0.757}$  \\ 
RGB$\rightarrow$LAB$\rightarrow$HSV & $25.32$ & ${0.761}$  & $26.31$ & ${0.760}$  \\ 
HSV$\rightarrow$LAB$\rightarrow$RGB & $26.53$ & ${0.787}$  & $26.67$ & ${0.769}$  \\ 
\hline
\end{tabularx}
}
\label{tab:color_block_permutation_results}
\end{table}

\begin{table}[t]
\centering
\caption{\small{Ablation study on the \textbf{Samsung S7} dataset~\cite{schwartz19} for the various terms in the CURL loss function. $rgb,no{-}cos$ is Equation~\ref{eq:rgbloss} \emph{without} the cosine term.}
}
\adjustbox{max width=\textwidth}{%
\begin{tabularx}{\linewidth}{ l | X X } 
\multicolumn{3}{c}{} \\
\textbf{\small{CURL Loss Terms}} & \textbf{\small{PSNR}} & \textbf{\small{SSIM}}  \\
\hline
\small{TED$+$CURL ($\mathcal{L}_{lab} {+} \mathcal{L}_{reg}$)} & \small{${26.19}$} & \small{${0.777}$} \\ 
\small{TED$+$CURL ($\mathcal{L}_{rgb} {+} \mathcal{L}_{reg}$)} & \small{${26.52}$}     &\small{${0.785}$}    \\ 
\small{TED$+$CURL ($\mathcal{L}_{rgb,no{-}cos} {+} \mathcal{L}_{reg}$)} & \small{${25.65}$}     &\small{${0.768}$}     \\ 
\small{TED$+$CURL ($\mathcal{L}_{hsv} {+} \mathcal{L}_{reg}$)} & \small{${{25.90}}$}     &\small{${0.777}$}    \\ 
\small{TED$+$CURL ($\mathcal{L}_{hsv} {+}\mathcal{L}_{rgb} {+} \mathcal{L}_{reg}$)} & \small{${26.61}$}     &\small{${{0.792}}$}   \\ 
\small{TED$+$CURL ($\mathcal{L}_{lab} {+} \mathcal{L}_{hsv} {+} \mathcal{L}_{reg}$)} & \small{${26.35}$}    &\small{${{0.781}}$}   \\ 
\small{TED$+$CURL ($\mathcal{L}_{lab} {+} \mathcal{L}_{rgb} {+} \mathcal{L}_{reg}$)} & \small{${25.88}$}     &\small{${{0.767}}$}   \\ 
\small{TED$+$CURL ($\mathcal{L}_{lab} {+} \mathcal{L}_{hsv} {+} \mathcal{L}_{rgb}$)} & \small{${26.32}$}&  \small{${{0.768}}$}   \\ 
\small{TED$+$CURL ($\mathcal{L}_{lab} {+} \mathcal{L}_{hsv} {+} \mathcal{L}_{rgb,no{-}cos} {+} \mathcal{L}_{reg}$)} & \small{${25.45}$}     &   \small{${{0.756}}$}  \\ 
\small{TED$+$CURL ($\mathcal{L}_{lab} {+} \mathcal{L}_{hsv} {+} \mathcal{L}_{rgb} {+} \mathcal{L}_{reg}$)} & \small{{$\textbf{27.09}$}}     &   \small{$\textbf{{0.793}}$}   \\ 
\hline
\end{tabularx}
}
\label{tab:loss_raw2rgb_ablation_study}
\end{table}

{\flushleft{\textbf{Colour Spaces of CURL:} Tables~\ref{tab:color_blocks_ablation}, \ref{tab:color_block_permutation_results} present ablation studies on the inclusion of individual colour spaces and on the ordering of the colour spaces within the retouching block. In Table~\ref{tab:color_blocks_ablation} we find that inclusion of all colour spaces (HSV, RGB, Lab) in the colour block leads to the best PSNR and SSIM, versus variants of the model that have only a single colour space (Figure~\ref{fig:colour_block_ablation}). Table~\ref{tab:color_block_permutation_results} suggests that the ordering of the colour spaces is important, with the highest image quality attained with a Lab, RGB and then HSV adjustment. \textbf{Loss function terms:} Equation~\ref{eq:difar_loss} presents the CURL loss function. We perform an ablation study in  Table~\ref{tab:loss_raw2rgb_ablation_study} on the various terms in the loss function.  The highest image quality is attained with all loss function terms, demonstrating the need to constrain each of the three colour spaces appropriately with a dedicated loss term. Interestingly we find that coupling the the $L_{1}$ RGB loss with a cosine distance term gives a significant boost in image quality (25.45$\rightarrow$27.09 dB, 0.756$\rightarrow$0.793 SSIM) compared to the using just the $L_{1}$ RGB loss term. In addition the regularization term ($\mathcal{L}_{reg}$) is important for highest image quality due to its role in constraining the flexibility of the neural retouching curves. }}

\subsection{Comparison to state-of-the-art methods}\label{sec:sota}

We evaluate our method against competitive baseline models. Specifically our main state-of-the-art baselines are (i) \textbf{U-Net:}~\cite{Chen2018DPE,Chen2018} we use a U-Net architecture without the MSCA-connections, and broadly following the design of the generator architecture of DPE~\cite{Chen2018DPE}. (ii) \textbf{DeepISP}~\cite{schwartz19}: we follow the architecture and experimental procedure of~\cite{schwartz19}. The thirty element RGB colour transformation matrix is  initialized using linear regression. (iii) \textbf{DPE:}~\cite{Chen2018DPE} we evaluate against the supervised (paired data) version of DPE. (iv) \textbf{DeepUPE}~\cite{wang19} (v) \textbf{HDRNet}~\cite{gharbi2017deep} (vi) \textbf{White-Box}~\cite{hu2018}.

\begin{table}[t] 
\centering
\small
\caption{\small\textbf{Medium-to-medium} exposure RAW to RGB mapping results on the held-out test images of the \textbf{Samsung S7 dataset}~\cite{schwartz19}.} 
\adjustbox{max width=\textwidth}{%
\begin{tabularx}{\linewidth}{ l | X X X X }
\textbf{Architecture} & \textbf{PSNR}$\uparrow$ & \textbf{SSIM}$\uparrow$ & \textbf{LPIPS}$\downarrow$ & \#~\textbf{Params} \\
\hline
TED$+$ CURL   & \textbf{27.04} & \textbf{0.794} & \textbf{0.320}   & 1.4 M\\ 
TED & ${26.56}$     &${0.781}$  & 0.339 &      1.3 M \\ 
\hline
\hline
U-Net~\cite{Ronneberger15}               & {25.90} & {0.783}  & 0.340    &  5.1 M  \\ 
DeepISP~\cite{schwartz19}                & {26.51} & {0.794} &  0.326  & 3.9 M \\ 
\hline
\end{tabularx}
}
\label{tab:s7_results1}
\end{table}

\begin{figure*}[t!]
\centering
\begin{tabular}{c@{}c@{}c@{}}
      \scalebox{0.85}{DeepISP (28.19 dB)} & 
      \scalebox{0.85}{\textbf{TED$+$CURL (29.37 dB)} } &
      \scalebox{0.85}{Groundtruth} \\
     \includegraphics[width=0.33\linewidth,scale=0.3]{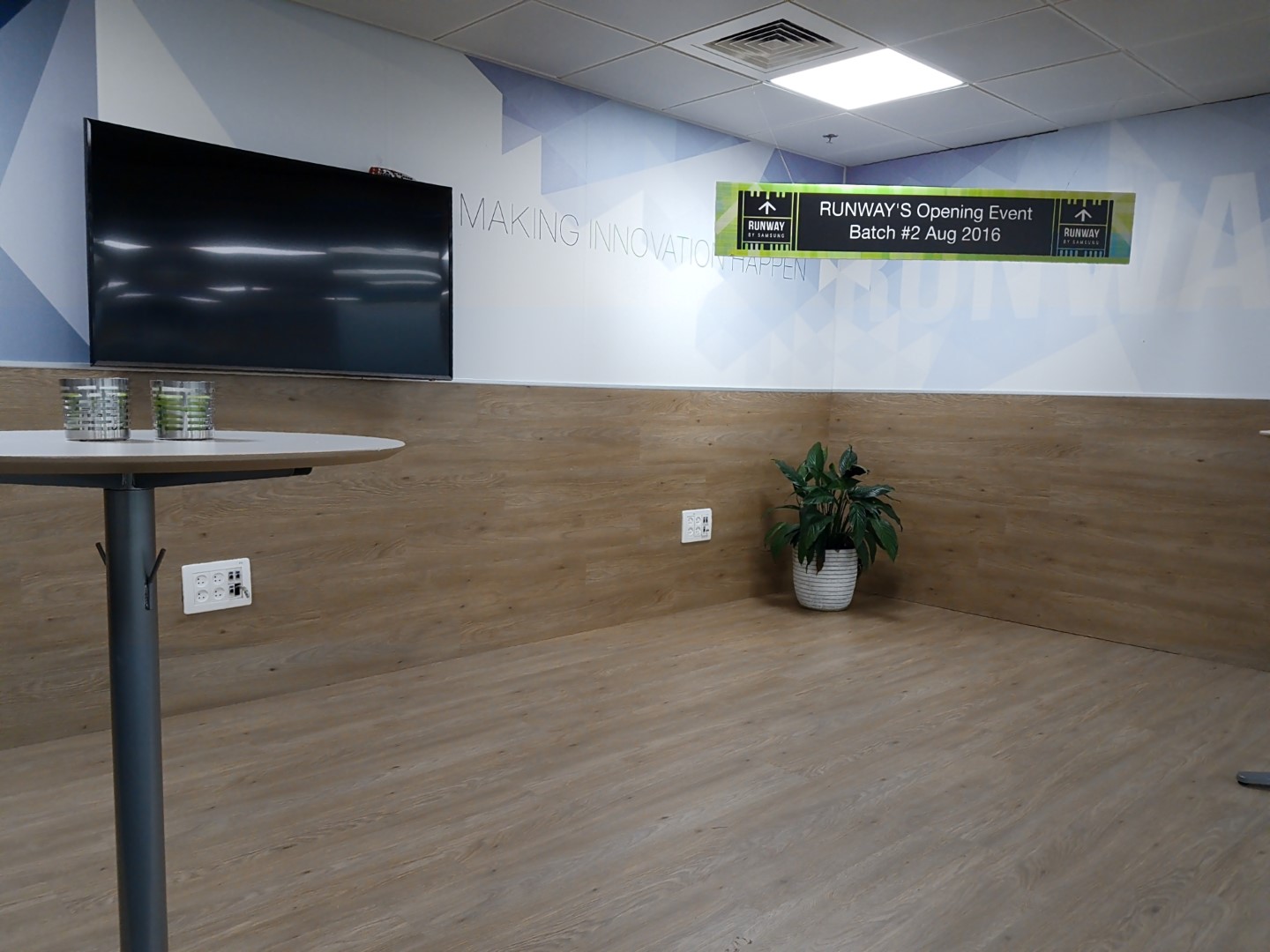} &
     \includegraphics[width=0.33\linewidth,scale=0.3]{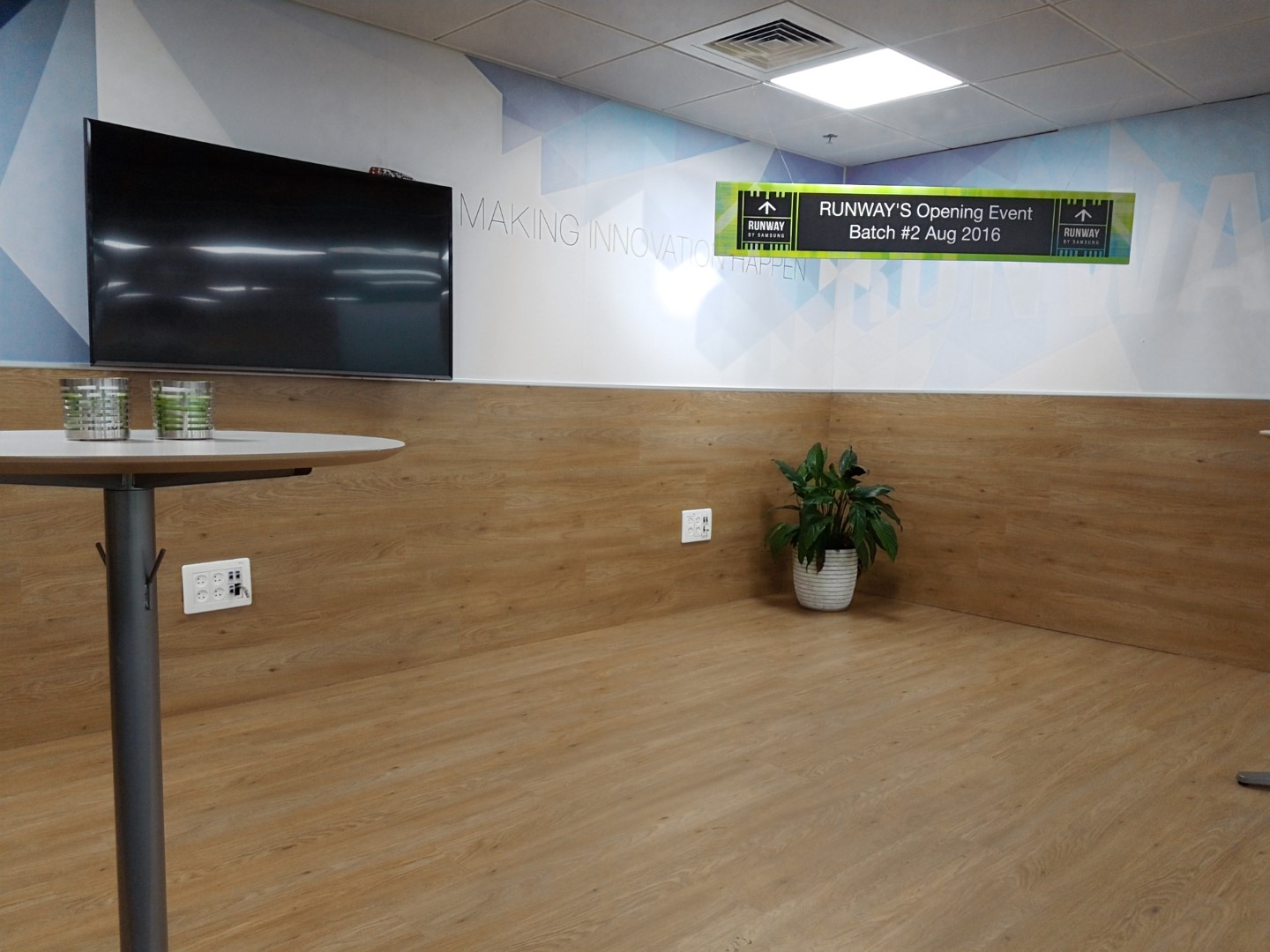} &
    \includegraphics[width=0.33\linewidth,scale=0.3]{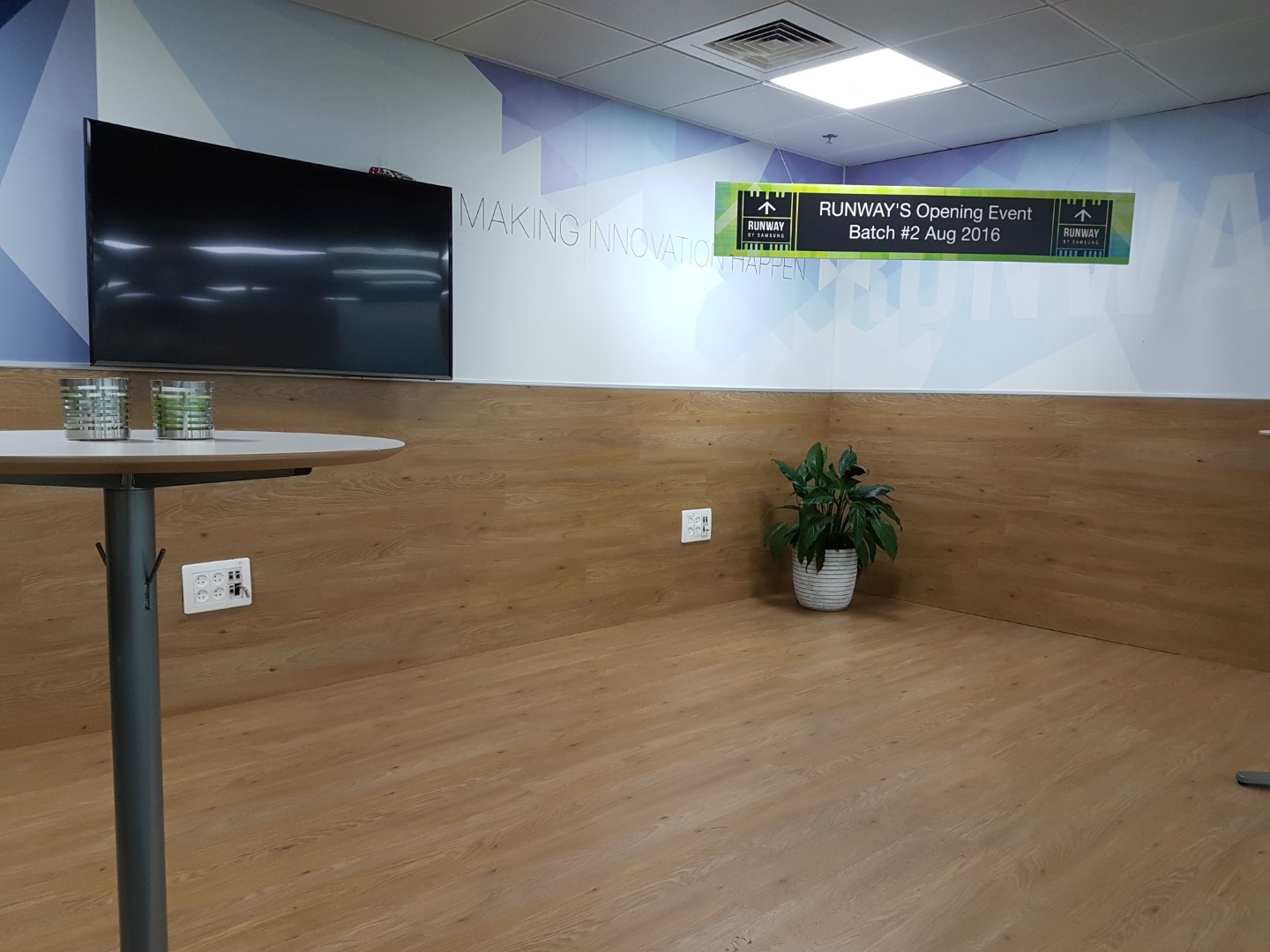} \\
 \end{tabular}
    \caption{Examples images produced by \textbf{DeepISP} and \textbf{TED$+$CURL} on the Samsung S7 \emph{Medium Exposure} dataset.}
    \label{fig:in_paper_samsung_medium_examples}
\end{figure*}

\begin{figure*}[t!]
\centering
\begin{tabular}{c@{}c@{}c@{}}
      \scalebox{0.85}{DeepUPE (16.85 dB)} & 
      \scalebox{0.85}{\textbf{TED$+$CURL (23.55 dB)} } &
      \scalebox{0.85}{Groundtruth} \\
     \includegraphics[width=0.33\linewidth,scale=0.3]{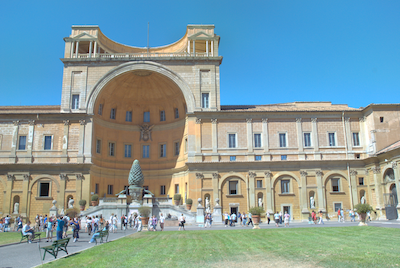} &
     \includegraphics[width=0.33\linewidth,scale=0.3]{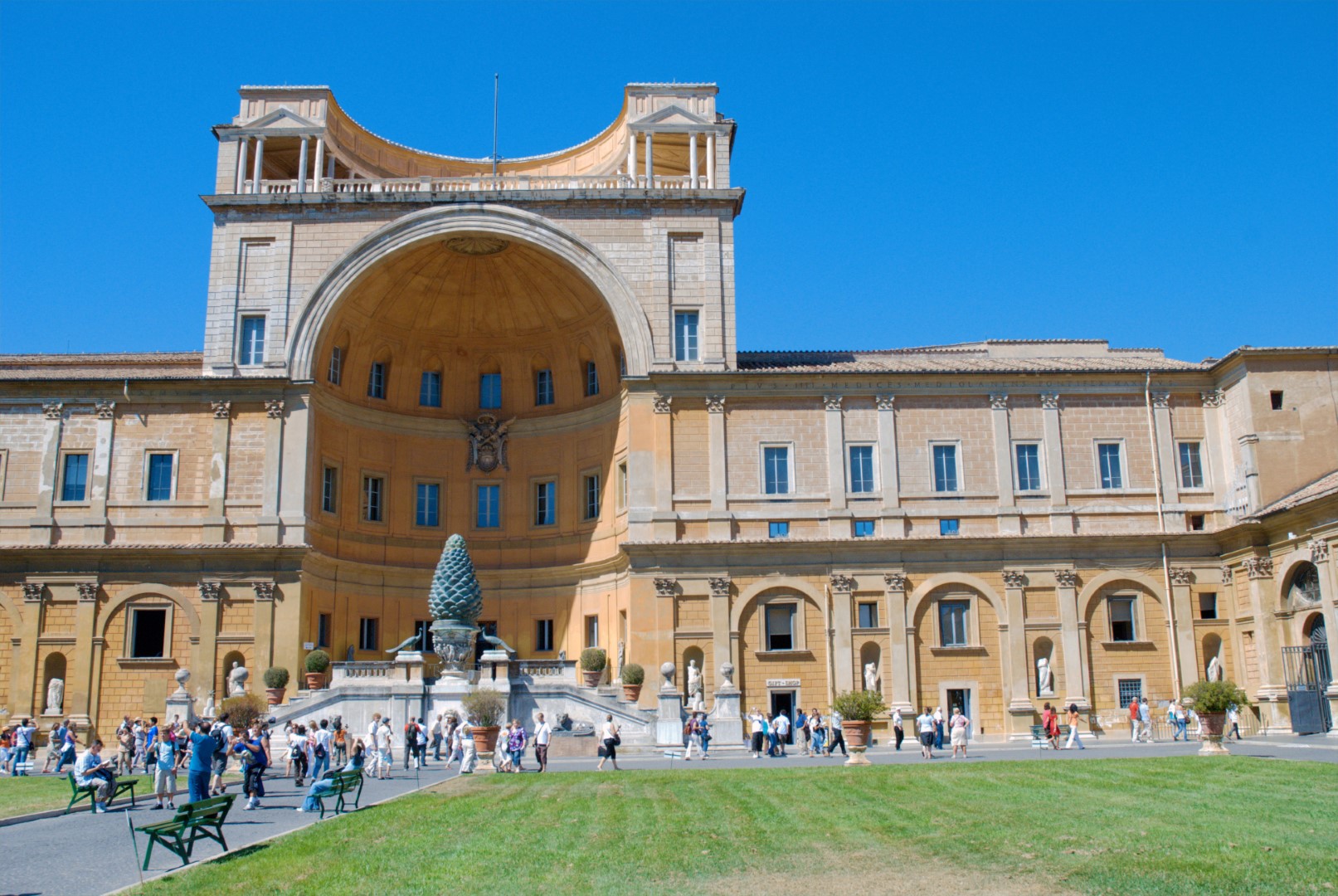} &
    \includegraphics[width=0.33\linewidth,scale=0.3]{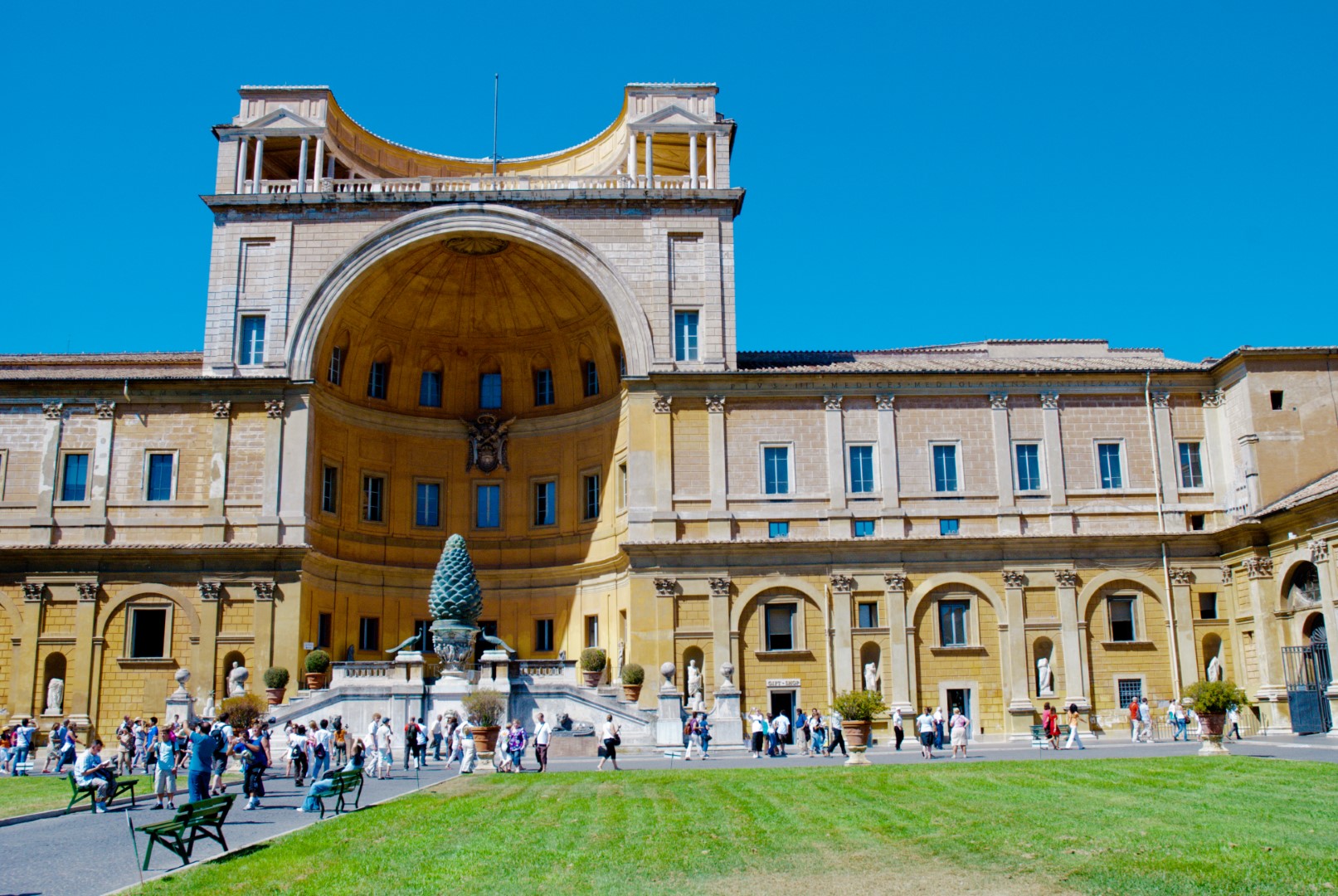} \\
\end{tabular}
    \caption{Examples images produced by \textbf{DeepUPE} and \textbf{TED$+$CURL} on the \emph{MIT-Adobe-UPE} dataset. }
    \label{fig:in_paper_adobe_upe_medium_examples}
\end{figure*}

{\flushleft{\textbf{Quantitative Comparison:}}}\quad \textbf{(i) Samsung S7 dataset:} We evaluate TED$+$CURL on the RAW-to-RGB mapping task for long-to-long exposure image pairs. TED$+$CURL is compared to U-Net~\cite{Ronneberger15} and DeepISP~\cite{schwartz19}. Results are presented in Table~\ref{tab:s7_results1}. TED$+$CURL produces higher-quality images for both exposure settings compared to U-Net and outperforms DeepISP in terms of PSNR and LPIPS metrics. This suggests TED$+$CURL is a competitive model for replacing the traditional RAW-to-RGB ISP pipeline. \textbf{(ii) MIT-Adobe5k (DPE):} Table~\ref{tab:dpe_results} presents the results on this dataset. TED$+$CURL uses ${\sim}2.5{\times}$ fewer parameters than DPE, yet is able to maintain the same SSIM score while boosting PSNR by $+0.17$dB. \textbf{(iii) MIT-Adobe5k (UPE):}  Table~\ref{tab:upe_results} demonstrates that CURL is competitive with DeepUPE on this challenging high-resolution dataset. TED$+$CURL obtains a substantial $+1.16$dB boost in PSNR, a reduction in LPIPS (lower is better) while retaining a competitive SSIM. \textbf{Visual Comparison:} results showing our method output in comparison to DeepISP and DeepUPE are shown in Figures~\ref{fig:in_paper_samsung_medium_examples}-\ref{fig:in_paper_adobe_upe_medium_examples}. CURL produces images with more pleasing colour and luminance compared to DeepISP and DeepUPE. Additional visual examples are presented in the supplementary material.

\begin{table}[t]
\small
\centering
\caption{\small Prediction quality of photographer C retouchings for the \textbf{MIT-Adobe 5K}~\cite{Chen2018DPE} test images. Results for other architectures are extracted from~\cite{Chen2018DPE}.} 
\adjustbox{max width=\textwidth}{%
\begin{tabularx}{\linewidth}{ l | X X X X}
\multicolumn{5}{c}{} \\
\textbf{Architecture} & \textbf{PSNR}$\uparrow$ & \textbf{SSIM}$\uparrow$  & \textbf{LPIPS}$\downarrow$ &  \#~\textbf{Params} \\
\hline
TED$+$CURL & \textbf{24.04} & \textbf{0.900} & \textbf{0.583} & 1.4 M \\ 
\hline
\hline
DPED~\cite{IgnatovICCV2017}   & ${21.76}$ & ${0.871}$ &-- &--  \\ 
8RESBLK~\cite{zhu2017unpaired,liu2017unsupervised}   & ${23.42}$ & ${0.875}$  & -- &--  \\ 
FCN~\cite{chen2017fast}    & ${20.66}$ & ${0.849}$  & -- & --  \\ 
CRN~\cite{chen2017photographic}   & ${22.38}$ & ${0.877}$  & -- &-- \\ 
U-Net~\cite{Ronneberger15}   & ${22.13}$ & ${0.879}$  & -- &-- \\ 
DPE~\cite{Chen2018DPE}   & 23.80 & 0.900 & 0.587 & 3.3 M \\ 
\hline
\end{tabularx}
}
\label{tab:dpe_results}
\end{table}

\begin{table}[t]
\small
\centering
\caption{\small Average PSNR, SSIM results for predicting the retouching of photographers on the 500 testing images from the \textbf{MIT-Adobe 5K dataset}. Dataset pre-processed according to the DeepUPE paper~\cite{wang19}. The PSNR and SSIM of other architectures are extracted from~\cite{wang19}.} 
\adjustbox{max width=\textwidth}{%
\begin{tabularx}{\linewidth}{ l | X X X X }
\multicolumn{5}{c}{} \\
\textbf{Architecture} & \textbf{PSNR}$\uparrow$ & \textbf{SSIM}$\uparrow$ & \textbf{LPIPS}$\downarrow$ & \#~\textbf{Params} \\
\hline
TED$+$CURL    & $\textbf{24.20}$ & ${0.880}$   & \textbf{0.108}  & 1.4 M \\ 
\hline
\hline
HDRNet~\cite{gharbi2017deep}   & ${21.96}$ & ${0.866}$  &-- &-- \\ 
DPE~\cite{Chen2018DPE}   & ${22.15}$ & ${0.850}$   &-- & 3.3 M\\ 
White-Box~\cite{hu2018}   & ${18.57}$ & ${0.701}$  & --& -- \\ 
Distort-and-Recover~\cite{park2018distort}    & 20.97 & 0.841  & -- & -- \\ 
DeepUPE~\cite{wang19}   & 23.04 & $\textbf{0.893}$      &  0.158  &1.0 M\\ 
\hline
\end{tabularx}
}
\label{tab:upe_results}
\end{table}

\section{Conclusions}
This paper introduced \emph{CURL} (\textbf{CUR}ve \textbf{L}ayers), a novel neural block for image enhancement. CURL takes inspiration from artists/photographers and retouches images based on global image adjustment curves. The retouching curves are learnt automatically during training to adjust image properties by exploiting image representation in three different colours spaces (CIELab, HSV, RGB). The adjustments applied by these curves is moderated by novel multi-colour space loss function. In our experimental evaluation a encoder/decoder backbone augmented with the pluggable CURL block significantly outperformed the state-of-the-art across a suite of benchmark datasets. Future research will investigate per-image adaptive ordering of adjustments in the CURL block.

\bibliographystyle{IEEEtran}
\bibliography{egbib}
%




\end{document}